%% file: paper.tex
\documentclass[pre,superscriptaddress,twocolumn]{revtex4-1}
\input{defs}
\def\lambda{\beta}

\begin{document}

\title{Entropy of labeled versus unlabeled networks}

\author{Jeremy Paton}
\affiliation{Department of Physics, Northeastern University, Boston, Massachusetts 02115, USA}
\affiliation{Network Science Institute, Northeastern University, Boston, Massachusetts 02115, USA}

\author{Harrison Hartle}
\affiliation{Network Science Institute, Northeastern University, Boston, Massachusetts 02115, USA}

\author{Huck Stepanyants}
\affiliation{Department of Physics, Northeastern University, Boston, Massachusetts 02115, USA}
\affiliation{Network Science Institute, Northeastern University, Boston, Massachusetts 02115, USA}

\author{Pim van der Hoorn}
\affiliation{Department of Mathematics and Computer Science, Eindhoven University of Technology, Postbus 513, 5600 MB Eindhoven, Netherlands}

\author{Dmitri Krioukov}
\affiliation{Department of Physics, Northeastern University, Boston, Massachusetts 02115, USA}
\affiliation{Network Science Institute, Northeastern University, Boston, Massachusetts 02115, USA}
\affiliation{Department of Mathematics, Northeastern University, Boston, Massachusetts 02115, USA}
\affiliation{Department of Electrical and Computer Engineering, Northeastern University, Boston, Massachusetts 02115, USA}

\begin{abstract}
The structure of a network is an unlabeled graph, yet graphs in most models of complex networks are labeled by meaningless random integers. Is the associated labeling noise always negligible, or can it overpower the network-structural signal? To address this question, we introduce and consider the sparse unlabeled versions of popular network models, and compare their entropy against the original labeled versions. We show that labeled and unlabeled \erdren graphs are entropically equivalent, even though their degree distributions are very different. The labeled and unlabeled versions of the configuration model may have different prefactors in their leading entropy terms, although this remains conjectural. Our main results are upper and lower bounds for the entropy of labeled and unlabeled one-dimensional random geometric graphs. We show that their unlabeled entropy is negligible in comparison with the labeled entropy. This means that in sparse networks the entropy of meaningless labeling may dominate the entropy of the network structure. The main implication of this result is that the common practice of using exchangeable models to reason about real-world networks with distinguishable nodes may introduce uncontrolled aberrations into conclusions made about these networks, suggesting a need for a thorough reexamination of the statistical foundations and key results of network science.
\end{abstract}

\maketitle

\section{Introduction}\label{sec:intro}

Networks are everywhere, and all of them are labeled. The labels of people in social networks are their names and all their other metadata, such as occupation, interests, place of living, and so on. Similarly, genes in gene regulatory networks, routers in the Internet, or countries in the world trade web, all have their unique meaningful names or identifiers. This labeling information is ignored if one is interested in the structure of a real-world network. Here, we assume that the \emph{structure} of a network is defined to be an \emph{unlabeled} graph, many visualizations of which can be found in textbooks, papers, or presentations in network science and graph theory.

Yet all of the popular network models used to study the structure of real-world networks are not unlabeled. Nodes in these models are actually labeled. However, there is a drastic difference between node labels in real-world networks and node labels in network models. Since network models are typically abstract mathematical models of random graphs, node labels in them cannot be as meaningful as the names of countries in the world, for instance. Node labels in network models come from arbitrary abstract sets of size~$n$, the network size. Without loss of generality, such label sets can be and usually are set to the set of integers from~$1$ to~$n$, denoted by~$[n]\denote\set{1,2, \ldots, n}$. Furthermore, since graphs in these models are typically random, so are labels in them. Any node in any \erdren graph of size~$100$ can have label~$99$, for instance, as opposed to the label~\textit{Bhutan} attached to an individual country in the world trade web.\\

Can the entropy coming from such random labeling of an unlabeled graph, the network structure, be safely ignored, or can it introduce non-negligible aberrations into the system that we have to account for in a nontrivial way?
In other words, since the meaningless random labels~$[n]$ are nothing but ``noise,'' assigned to an unlabeled graph uniformly at random out of the $n!$ permutations, then can it be the case that this noise statistically dominates the randomness associated with the network structure, an unlabeled graph?

The principled way to address this question is to compare the leading terms of network entropy in the labeled and unlabeled cases. If the former dominates the latter, then indeed the noise overpowers the signal. The other reason to focus on entropy is that entropy is one of the most important properties of a network model, a central player in the definitions of the unbiased null models of networks~\cite{park2004statistical, bianconi2008entropy, bianconi2008entropies, bianconi2009entropy, anand2009entropy, anand2011shannon, garlaschelli2008maximum, garlaschelli2009generalized, squartini2011analytical, mastrandrea2014enhanced, squartini2015unbiased, zuev2015exponential, hoorn2018sparse}, network ensemble equivalence~\cite{anand2009entropy, anand2010gibbs, squartini2015breaking, garlaschelli2017ensemble}, network typicality~\cite{shannon1948mathematical, cover2005elements}, and many other fundamental matters~\cite{bianconi2009assessing, zhao2011entropy_rate, peixoto2012entropy, peixoto2013parsimonious, anand2014entropy, racz2017basic, radicchi2018uncertainty, radicchi2020classical, bianconi2022statistical, bianconi2022grand, cimini2019statistical, coutrot2022entropy}. Yet our main motivation are null models, which are network models that maximize network entropy under various network-structural constraints.\\

Here we show that maximizing the entropy of the naked network structure represented by an unlabeled graph, and maximizing the entropy of this structure dressed in random labels represented by a labeled graph, may lead to very different outcomes in sparse networks. The entropy of meaningless random labeling may be a dominating factor, so it gets maximized, instead of the intended maximization of network-structural entropy. 

In what follows, we first recall in Section~\ref{sec:background} the key differences between labeled and unlabeled graphs, and explain why in most cases---essentially in all the cases that deal with maximum-entropy null models of real networks---one should be interested in unlabeled graph models, versus their well-known labeled counterparts. In a nutshell, this is because labels in real networks are ``glued'' to individual nodes, like \textit{Bhutan} to the country, resulting in one network structure, as opposed to labeled network models where this structure is repeated as many times as the number of graph isomorphisms, i.e., the number of label permutations leading to a different but isomorphic labeled graph. As a consequence, labeled models are biased towards more asymmetric graphs, with larger isomorphism classes, compared to the unbiased unlabeled models of the network structure with the same sufficient statistics.

In Section~\ref{sec:ERCM}, we then consider the two most basic illustrative examples: (1)~the ``harmonic oscillator'' of network models---the \erdren random graphs~(ER), and (2)~the configuration model~(CM). In the ER case, we consider the unlabeled versions of both microcanonical $\cG_{n,m}$ and canonical $\cG_{n,p}$ labeled ER graphs, denoting these unlabeled ER models by $\cU_{n,m}$ and $\cU_{n,p}$. Somewhat shockingly, the unlabeled canonical ER graphs $\cU_{n,p}$ have neither been considered nor even properly defined before. It is known, however, that even such a basic property as the degree distribution is very different in sparse microcanonical labeled $\cG_{n,m}$ versus unlabeled $\cU_{n,m}$ ER graphs---here by \emph{sparse graphs} we mean graphs with constant average degree~$\bar{k}=2m/n$. Notwithstanding these differences, we show that the leading term of entropy of both $\cG_{n,m}$ and $\cU_{n,m}$ is surprisingly the same~$(\bak/2) n \log n$. The subleading entropy terms are different, however. The leading term of entropy of the unlabeled microcanonical CM with scale-free degree distributions is unknown but is not excluded to be the same as in the labeled case, albeit with a different prefactor, $(\bak/2-1) n \log n$. The unlabeled canonical CM has never been mentioned before either, so we define it in Section~\ref{sec:ERCM} as well.

Our main results are in Section~\ref{sec:RGG}. They are tight lower and upper bounds for the entropy of unlabeled and labeled one-dimensional random geometric graphs~(RGGs). The calculation of RGG entropy is an important longstanding problem that has seen only limited progress as it has been considered intractable~\cite{coon2018entropy, badiu2018distribution, bubeck2016testing, liu2021phase}. We develop a powerful technique, rooted in the labeled-unlabeled delineation, that allows us to show that the leading terms of the entropy of sparse unlabeled and labeled RGGs are different, $\lesssim n$ versus $\sim n \log n$, respectively, Table~\ref{tab:summary}. This disconcerting result implies that the entropy of labeled graphs is dominated by the entropy of the meaningless labeling noise, rather than by the entropy of the network structure. 

\begin{table}
  \centering
  \begin{tabular}{|l|l|l|l|}
     \hline
     Entropy & ER & CM & RGG \\ \hline 
     Labeled & $\approx(\bak/2)n\log n$ & $\approx(\bak/2)n\log n$ & $\sim n\log n$ \\
     Unlabeled & $\approx(\bak/2)n\log n$ & $\approx(\bak/2-1)n\log n$ & $\lesssim n$ \\
     \hline
   \end{tabular}
  \caption{The leading terms of labeled and unlabeled entropies of the sparse microcanonical \erdren random graphs (ER), configuration model with scale-free degree sequences (CM), and one-dimensional random geometric graphs (RGG). The scaling of unlabeled CM entropy is conjectural.
  }\label{tab:summary}
\end{table}

The entropic equivalence is a necessary but not sufficient condition for the ensemble equivalence~\cite{anand2009entropy, anand2010gibbs, squartini2015breaking, garlaschelli2017ensemble}. Our RGG result thus says that labeled and unlabeled RGGs are statistically very different. Yet even if the leading entropy terms are the same, as in labeled and unlabeled ER graphs, the ensembles can still be very different, up to the point that their degree distributions can be very different. The main overall conclusions are then that unlabeled network models may behave very differently from their labeled counterparts, so any predictions concerning the network structure based on labeled models may lead to potentially misleading or statistically incorrect outcomes. These and other implications and challenges are discussed in the concluding Section~\ref{sec:conclusion}.\\

\noindent{\bfseries Notations and conventions.} In what follows, the symbols `$\ll$', `$\sim$', `$\approx$', and `$\gg$' in $a_n$ `$\ast$' $b_n$ mean that $c=\lim_{n\to\infty}a_n/b_n$ is $c=0$, $0<c<\infty$, $c=1$, and $c=\infty$, respectively. We call networks sparse or dense if their (expected) average degree is $\bak\sim1$ or $\bak\sim n$. All networks are sparse below, unless mentioned otherwise. The adjunctive \emph{with high probability} is implied where needed.

\section{Labeled vs.\ unlabeled networks}\label{sec:background}

\begin{figure}
  \centerline{\includegraphics[width=\linewidth]{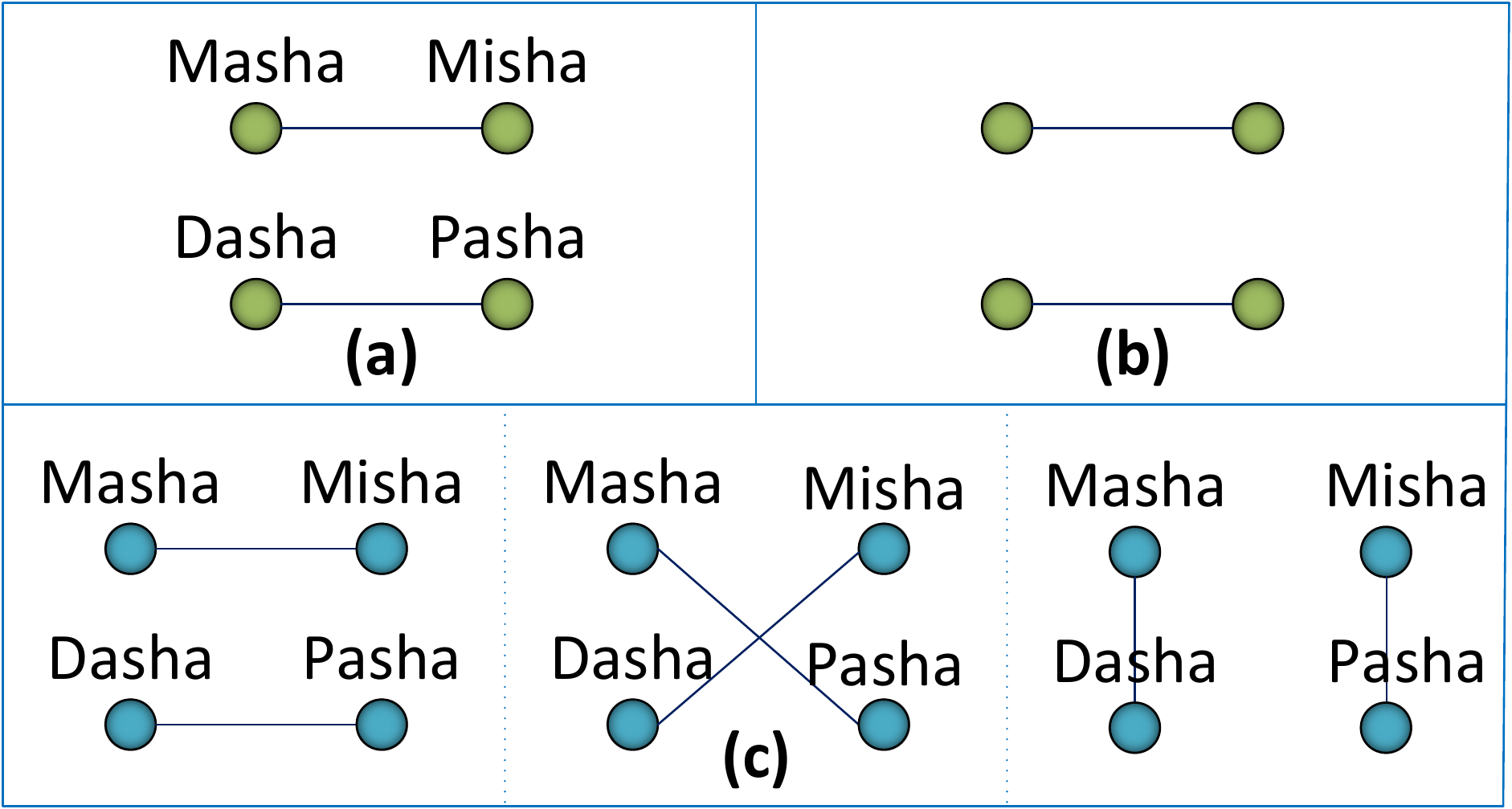}}
  \caption{{\bf A network of love:} a ``real-world'' network~(a) and its unlabeled~(b) and exchangeably labeled~(c) versions.}
  \label{fig:love}
\end{figure}

Consider a ``real-world'' love network among the four people in Fig.~\ref{fig:love}(a). Masha loves Misha while Dasha loves Pasha. The nodes are labeled by lovers' names, so the graph is labeled. What permutations of labels are allowed in this network? Clearly, we can only swap Masha with Misha and/or Dasha with Pasha. These swaps are called graph \emph{automorphisms}: before and after the swap the network is the same labeled graph. No other permutation of labels is allowed. We cannot swap Masha with Dasha or with Pasha, for instance, because such label permutations lead to different labeled graphs, with different love stories in the real life. Labels in real-world networks are thus ``glued'' to nodes since nodes are \emph{distinguishable} entities.

It is now critical to recognize that because of this gluing---or more formally, since no non-automorphism label permutations are allowed in real networks---we are essentially dealing with \emph{unlabeled} networks. Indeed, if we are not concerned who exactly loves whom exactly---that is, if we are interested only in the network \emph{structure}---then the network structure in the example is a pair of couples, represented by the \emph{unlabeled} graph in Fig.~\ref{fig:love}(b). If we \emph{are} interested in who loves whom, then again it is just \emph{one labeled} graph in Fig.~\ref{fig:love}(a). In either case, we are dealing with just \emph{one} graph, either labeled or unlabeled, and not with the \emph{three} isomorphic graphs labeled in all the three possible ways in Fig.~\ref{fig:love}(c). (Two labeled graphs are called \emph{isomorphic} if they are the same unlabeled graph.) Only \emph{one} of these three labeled graphs reflects reality; the other two are ``noise.''

Unlabeled graphs can but do not have to be considered as isomorphism classes of labeled graphs. In fact, representing unlabeled graphs as isomorphism classes of labeled graphs can be confusing. The easiest way to understand unlabeled graphs is via their enumeration~\cite{harary1973graphical}, a simple example of which we consider next.

\begin{figure}
  \centerline{\includegraphics[width=\linewidth]{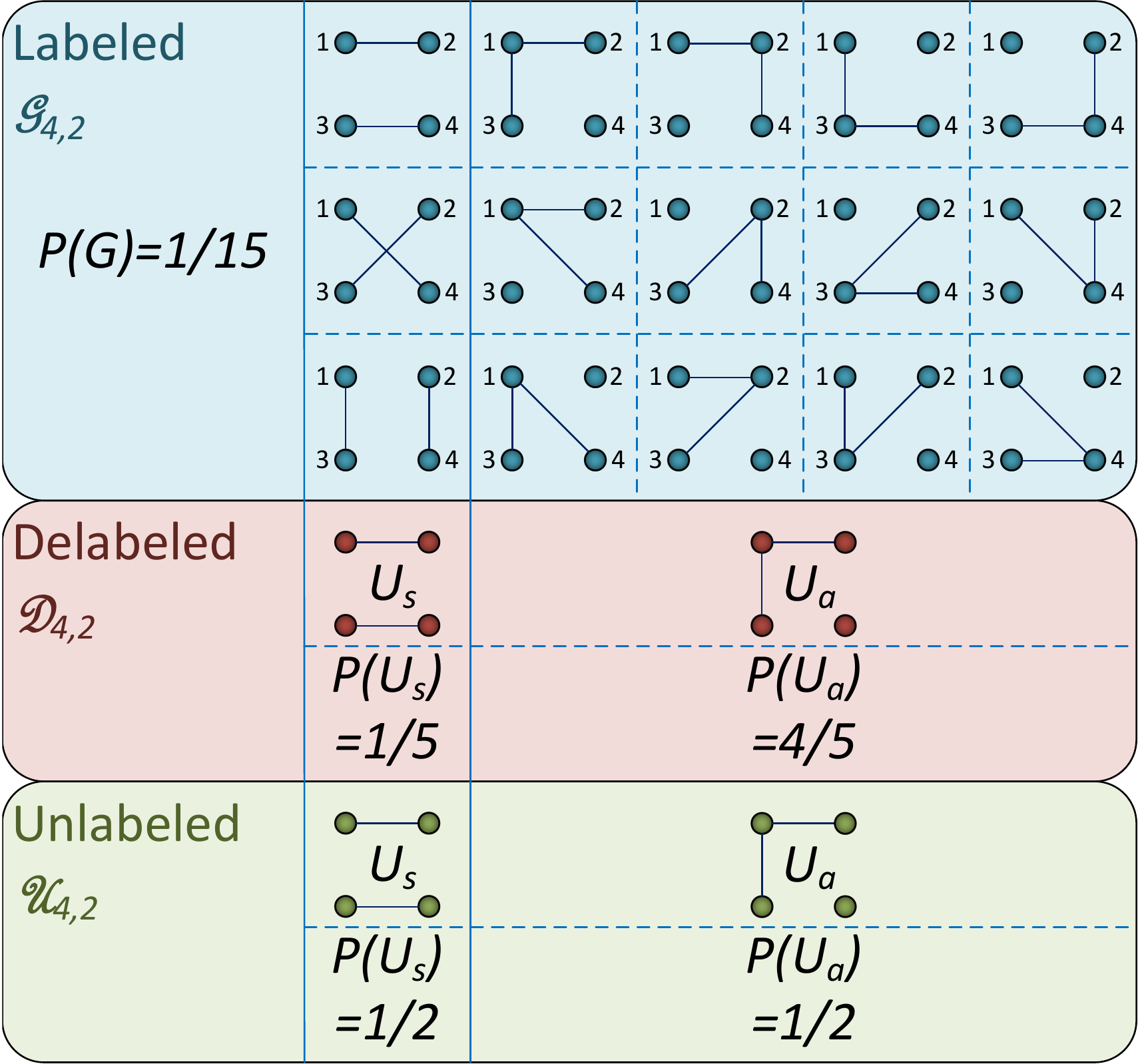}}
  \caption{{\bf The labeled, delabeled, and unlabeled microcanonical \erdren random graphs of size~$n=4$ with $m=2$ edges.} The probabilities of all labeled graphs~$G$ are the same $P(G)=1/15$ in $\cG_{n,m}$, while the probabilities of the symmetric and asymmetric unlabeled graphs $U_s$ and $U_a$ are different in the delabeled and unlabeled models~$\cD_{n,m}$ and~$\cU_{n,m}$.}
  \label{fig:erm}
\end{figure}

Suppose we are to formulate the statistically correct null model of networks with $n=4$ nodes and $m=2$ edges, as in our example with Misha, Masha, Pasha, and Dasha. By \emph{statistically correct} models we mean here the unbiased models that maximize entropy subject to given constraints~\cite{park2004statistical, bianconi2008entropy, bianconi2008entropies, bianconi2009entropy, anand2009entropy, anand2011shannon, garlaschelli2008maximum, garlaschelli2009generalized, squartini2011analytical, mastrandrea2014enhanced, squartini2015unbiased, zuev2015exponential, hoorn2018sparse}. In our example, these constraints are $n=4$ and $m=2$, so the correct entropy-maximizing null models are \emph{defined} by the uniform distributions over the space of all graphs with $n=4$ nodes and $m=2$ edges. However, this space is very different for unlabeled versus labeled graphs. There are only two unlabeled graphs with $4$ nodes and $2$ edges, while there are $15$ labeled ones, all shown in Fig.~\ref{fig:erm}. The uniform distribution $P(G)=1/15$ over the $15$ labeled graphs~$G$ is the familiar \erdren model~$\cG_{n,m}$ with $n=4$ and $m=2$, while the uniform distribution $P(U)=1/2$ over the two unlabeled graphs~$U$ is its virtually unknown unlabeled counterpart~$\cU_{n,m}$. Which one, $\cG_{n,m}$ or~$\cU_{n,m}$, are we supposed to work with in applications to real networks?

The answer to this question depends on whether the nodes in a real network are distinguishable or indistinguishable. The most crucial difference between the considered ``real'' love network and the labeled ER~$\cG_{4,2}$ is that the nodes in the love network are distinguishable, while they are indistinguishable in~$\cG_{4,2}$. Indeed, the label set in~$\cG_{4,2}$ is not \{Misha, Masha, Pasha, Dasha\} but \{1, 2, 3, 4\}, and the model is exchangeable. A model of labeled graphs is called \emph{exchangeable} if the probabilities of any two isomorphic graphs in the model are the same. Exchangeability is thus a statistical formalization of the idea that node labels ``do not matter'' and can be permuted arbitrarily. In other words, \emph{exchangeability is a formalization of statistical indistinguishability}.

Therefore, the very common practice of applications of \emph{exchangeable} models of labeled graphs to real networks with \emph{distinguishable} nodes is statistically questionable. Either \emph{nonexchangeable} models, in which node labels ``do matter and are glued to nodes,'' must be used for such networks, or---as far as null models of network structure are concerned---the statistically correct null models of such networks must be models of \emph{unlabeled} graphs. In either case, we are dealing not with three graphs in Fig.~\ref{fig:love}(c), but with one graph, either in Fig.~\ref{fig:love}(a) or in Fig.~\ref{fig:love}(b).

The other way around, if nodes in a real-world network are \emph{indistinguishable}, such as atoms in material networks~\cite{papadopoulos2018network}, then the statistically correct null models of such networks must be \emph{exchangeable} models of \emph{labeled} graphs. Indeed, different atoms are different atoms, but since they are indistinguishable, any permutation of individual atoms in a particular configuration is equally good, a typical situation in statistical physics~\cite{kittel1980thermal}, perhaps the most vivid illustration of which is the infamous Gibbs paradox~\cite{ford2013statistical}. However, such situations in the science of real-world complex networks are rare exclusions rather than a rule, since in a vast majority of real networks, nodes are distinguishable.

But does it really matter which models to use, labeled or unlabeled, as they may be equivalent in some way? The key message of this paper is that it really does matter what models we deal with as the labeled and unlabeled versions of the same model can be very nonequivalent. We already see clear signs of this in our toy example with $\cG_{4,2}$ and $\cU_{4,2}$ in Fig.~\ref{fig:erm}. The two models are clearly very different in many respects. For instance, the probability of the asymmetric $U_a$ scenario in Fig.~\ref{fig:erm} with one lucky person loving two others, while the unlucky fourth is left loveless, is~$80\%$ in the labeled~$\cG_{4,2}$, while it is only~$50\%$ in the unlabeled~$\cU_{4,2}$. That is, the two models give different predictions concerning the likelihoods of different ``love scenarios in the real life.''

\begin{figure}
  \centerline{\includegraphics[width=.75\linewidth]{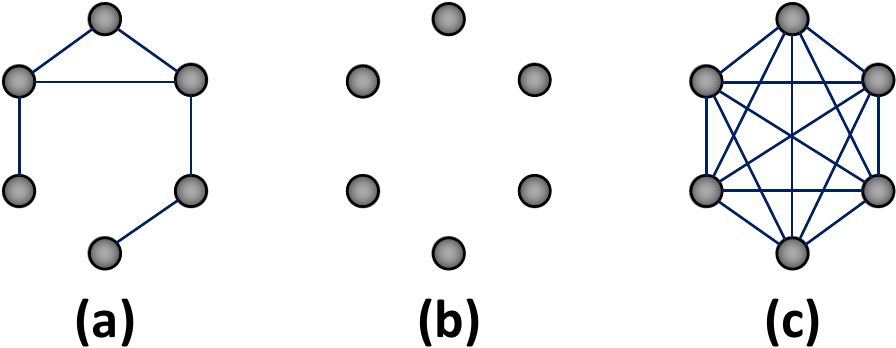}}
  \caption{{\bf The graph symmetry extremes:} the smallest and sparsest fully asymmetric graph~(a) and the fully symmetric empty~(b) and complete~(c) graphs of the same size.}
  \label{fig:66}
\end{figure}

The last observation illustrates the role of the symmetry of a graph in analyzing the statistical differences between labeled and unlabeled graph models. The graph~$U_a$ in Fig.~\ref{fig:erm} is ``more asymmetric'' than the pair of couples~$U_s$ because there are more label permutations on $U_a$'s labeled version that lead to different labeled graphs. In graph theory, a graph is called \emph{asymmetric} if any permutation of its labels is not an automorphism, i.e.\ if any label permutation leads to a different labeled graph. The graph is called \emph{symmetric} otherwise. The smallest asymmetric graphs are of size~$6$, and there are $8$ such graphs. The single one with the fewest edges~($6$) is shown in Fig.~\ref{fig:66}(a). In~$\cU_{6,6}$, it is just one unlabeled graph, but in~$\cG_{6,6}$, it corresponds to $6!=720$ labeled graphs, in stark contrast with the fully symmetric empty or complete graphs, Fig.~\ref{fig:66}(b,c), represented by only one graph, either unlabeled or labeled, since any label permutation of an empty or complete graph is an automorphism.

These observations are directly related to the common confusion between \emph{unlabeled} graph models and \emph{delabeled} graph models~\cite{choi2012compression, kontoyiannis2020compression}. A delabeled graph model starts with a labeled graph model, generates a labeled graph, and then simply removes the node labels in it. The result is a model of unlabeled graphs, in which more asymmetric graphs attract higher probability masses. In our example above, the delabeled ER model~$\cD_{n,m}$ with $n=4$ nodes and $m=2$ edges is the probability distribution on the two unlabeled graphs in Fig.~\ref{fig:erm}, which assigns the probability of~$12/15=80\%$ to the freaky asymmetric scenario~$U_a$, and only $3/15=20\%$ to the more conventional pair of couples~$U_s$, as opposed to the unlabeled~$\cU_{n,m}$, which says the two scenarios are equally likely, the probability of each is~$1/2=50\%$. In other words, if you generate a random $\cG_{n,m}$ graph (e.g.\ by placing $m$ edges randomly among ${n\choose2}$ node pairs), and then consider the generated graph as unlabeled, then you have sampled a random unlabeled graph \emph{not} from the unlabeled model~$\cU_{n,m}$, but from the delabeled model~$\cD_{n,m}$. This shows that as far as the probability of the \emph{network structure} is concerned, a delabeled graph model is equivalent to its labeled source, while both are different from the corresponding unlabeled model as Fig.~\ref{fig:erm} illustrates.\\ 

The key points of this illustrative section motivating what follows are:
\begin{itemize}
  \item exchangeability is indistinguishability of random variables;
  \item therefore, models of real networks with distinguishable nodes can be either nonexchangeable or unlabeled;
  \item the correct null models of the structure of real networks with distinguishable nodes must be models of unlabeled graphs;
  \item the statistical properties of unlabeled graph models can be very different from their labeled/delabeled counterparts;
  \item compared to their labeled/delabeled counterparts, unlabeled graph models have been studied much more poorly because they are much more difficult to deal with, see Section~\ref{sec:conclusion}.
\end{itemize}

\section{\erdren graphs and configuration model}\label{sec:ERCM}

\subsection{\erdren (ER) graphs}

As discussed in the previous section, the microcanonical labeled and unlabeled ER graph models~$\cG_{n,m}$ and~$\cU_{n,m}$ are \emph{defined by} the entropy-maximizing uniform probability distributions $P(G)=1/\size{\cG_{n,m}}$ and $P(U)=1/\size{\cU_{n,m}}$ over all labeled graphs~$G\in\cG_{n,m}$ and, respectively, unlabeled graphs~$U\in\cU_{n,m}$ with $n$ nodes and $m$ edges. While the number of labeled graphs with $n$ nodes and $m$ edges is exactly $\size{\cG_{n,m}}={N\choose m}$ where $N={n \choose 2}$, the number of unlabeled graphs with $n$ nodes and $m$ edges $\size{\cU_{n,m}}$ is known only asymptotically for large graphs~\cite{wright1974graphs}.

The conjugated canonical versions of~$\cG_{n,m}$ and~$\cU_{n,m}$ are~$\cG_{n,p}$ and~$\cU_{n,p}$. These are the maximum-entropy labeled and unlabeled graphs of size~$n$ in which the number of links is not fixed exactly to~$m$; instead the average number of links~$\bam$ is fixed to $pN$, or equivalently, the average graph density~$\bad=\bam/N$ is fixed to~$p$.
While $\cG_{n,p}$ is as well studied as $\cG_{n,m}$, the unlabeled canonical ER graphs $\cU_{n,p}$ has never been considered before, so we \emph{define} them next, after recalling the basic entropic facts about~$\cG_{n,p}$.

\subsubsection{Unlabeled canonical ER graphs~$\cU_{n,p}$}

As is well known, a $\cG_{n,p}$ graph~$G$ can be generated by linking all pairs of labeled nodes independently with probability~$p$. The resulting probability to generate graph~$G$ in the model is
\beq\label{eq:P(G)Gnp}
P(G)=p^{m(G)}(1-p)^{N-m(G)},
\eeq
where $m(G)$ is the number of edges in~$G$. This probability distribution is the canonical entropy-maximizing Gibbs (a.k.a.\ exponential family) distribution, since it can be rewritten in the Gibbs form~\cite{park2004statistical}
\beq
P(G)=\frac{\exp[\lambda m(G)]}{Z},
\eeq
where the partition function involves the summation over all labeled graphs~$\cG_n$ of size~$n$,
\beq
Z=\sum_{G\in\cG_n}\exp[\lambda m(G)],
\eeq
which can be shown simplifies to
\beq\label{eq:ZGnp}
Z=(e^\lambda+1)^N.
\eeq
The inverse temperature parameter~$\beta\in\R$ is related to~$p\in[0,1]$ via
\beq\label{eq:p(beta)Gnp}
p=\frac{1}{e^{-\lambda}+1},
\eeq
which is the solution of the standard free energy equation
\beq\label{eq:FGnp}
\frac{\partial\log Z}{\partial\beta}=\sum_{G\in\cG_n}m(G)P(G)=\bam=pN.
\eeq

The unlabeled ER graphs~$U$ in~$\cU_{n,p}$ are thus \emph{defined} by the entropy-maximizing probability distribution of the same Gibbs form,
\beq
P(U)=\frac{\exp[\lambda m(U)]}{Z},
\eeq
except that the graphs are unlabeled, so the partition function involves the summation not over all the $n$-sized \emph{labeled} graphs~$\cG_n$, but over the much smaller but also much more intractable space~$\cU_n$ of all the \emph{unlabeled} graphs of size~$n$,
\beq\label{eq:ZUnp}
Z=\sum_{U\in\cU_n}\exp[\lambda m(U)].
\eeq 
Unfortunately, this sum does not in general simplify to anything as nice-looking as~\eqref{eq:ZGnp}. As a consequence, there is no nice-looking analogy of~\eqref{eq:P(G)Gnp} for~$P(U)$, which, among many other things, implies that the $\cU_{n,p}$ graphs \emph{cannot} be generated by placing edges independently with probability~$p$ among $N$ unlabeled node pairs. By doing so, you generate an unlabeled graph not from the unlabeled model~$\cU_{n,p}$, but from the delabeled one~$\cD_{n,p}$. The free energy equation~\eqref{eq:FGnp} linking~$\beta$ to~$p$ holds,
\beq\label{eq:FUnp}
\frac{\partial\log Z}{\partial\beta}=\sum_{U\in\cU_n}m(U)P(U)=\bam=pN,
\eeq
but does not lead to anything as simple as~\eqref{eq:p(beta)Gnp}.
Yet we can show that the solution of~\eqref{eq:FUnp} exists and is unique for any $n,p$.

\begin{figure}
  \centerline{\includegraphics[width=\linewidth]{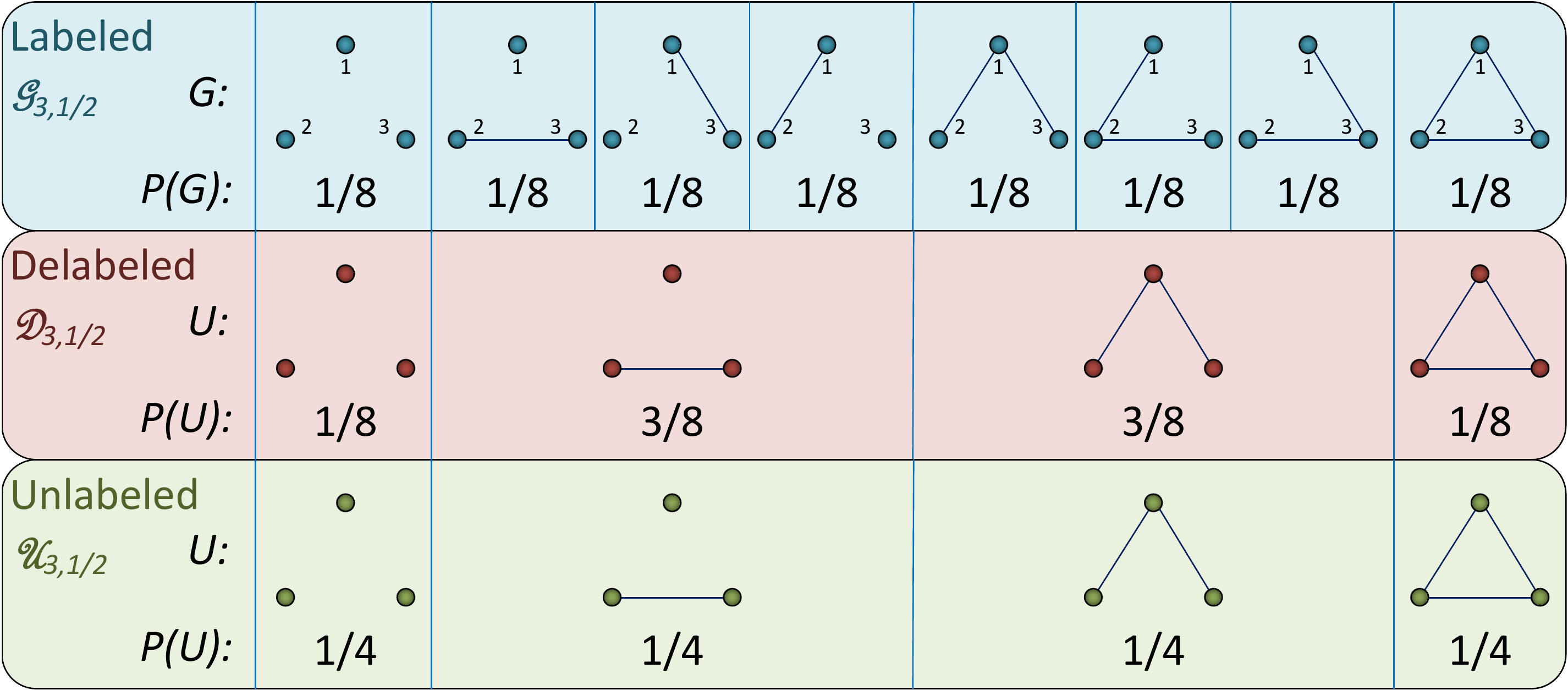}}
  \caption{{\bf The labeled, delabeled, and unlabeled canonical \erdren random graphs of size~$n=3$ and average density $p=1/2$.} The probabilities of all graphs in the models are shown at the bottom of each row. The entropies of the three models are $S_L=3$, $S_U=2$, and $S_D\approx1.81$ bits, reflecting the general inequality $S_D \leq S_U \leq S_L$.}
  \label{fig:erp}
\end{figure}

Using graph complementarity arguments, we can also show that the solution of~\eqref{eq:FUnp} with $p=1/2$ yields $\lambda=0$ (infinite temperature) for any $n$, resulting in the uniform distribution over all the unlabeled graphs~$\cU_n$ of size~$n$ as Fig.~\ref{fig:erp} illustrates for the simplest nontrivial case $n=3$. In $\cU_{3,p}$, the direct evaluation of~\eqref{eq:ZUnp} yields the partition function
\beq
Z=\frac{e^{4\lambda}-1}{e^{\lambda}-1},
\eeq
so the probability of the four unlabeled graphs of size~$3$ with $m=0,1,2,3$ edges is~$P(m)=e^{m\beta}/Z$, while~\eqref{eq:FUnp} leads to
\beq
p=\frac{1}{3}\left[\frac{1}{e^{-\lambda}+1}+\tanh\lambda+1\right].
\eeq
Figure~\ref{fig:erp} also illustrates the key statistical differences between the labeled and unlabeled ER graphs. For instance, the probability to generate a graph with $m=0,1,2,3$ edges in the labeled and delabeled cases with $\beta=0$ is given by the binomial distribution $1/8, 3/8, 3/8, 1/8$, while in the unlabeled case this probability is uniform, $1/4$ for any $m$. The statistical similarities between the delabeled and labeled graphs, both different from the unlabeled ones, are similar to those in the microcanonical case in Fig.~\ref{fig:erm}.

\begin{figure}
	\centerline{\includegraphics[width=.75\linewidth]{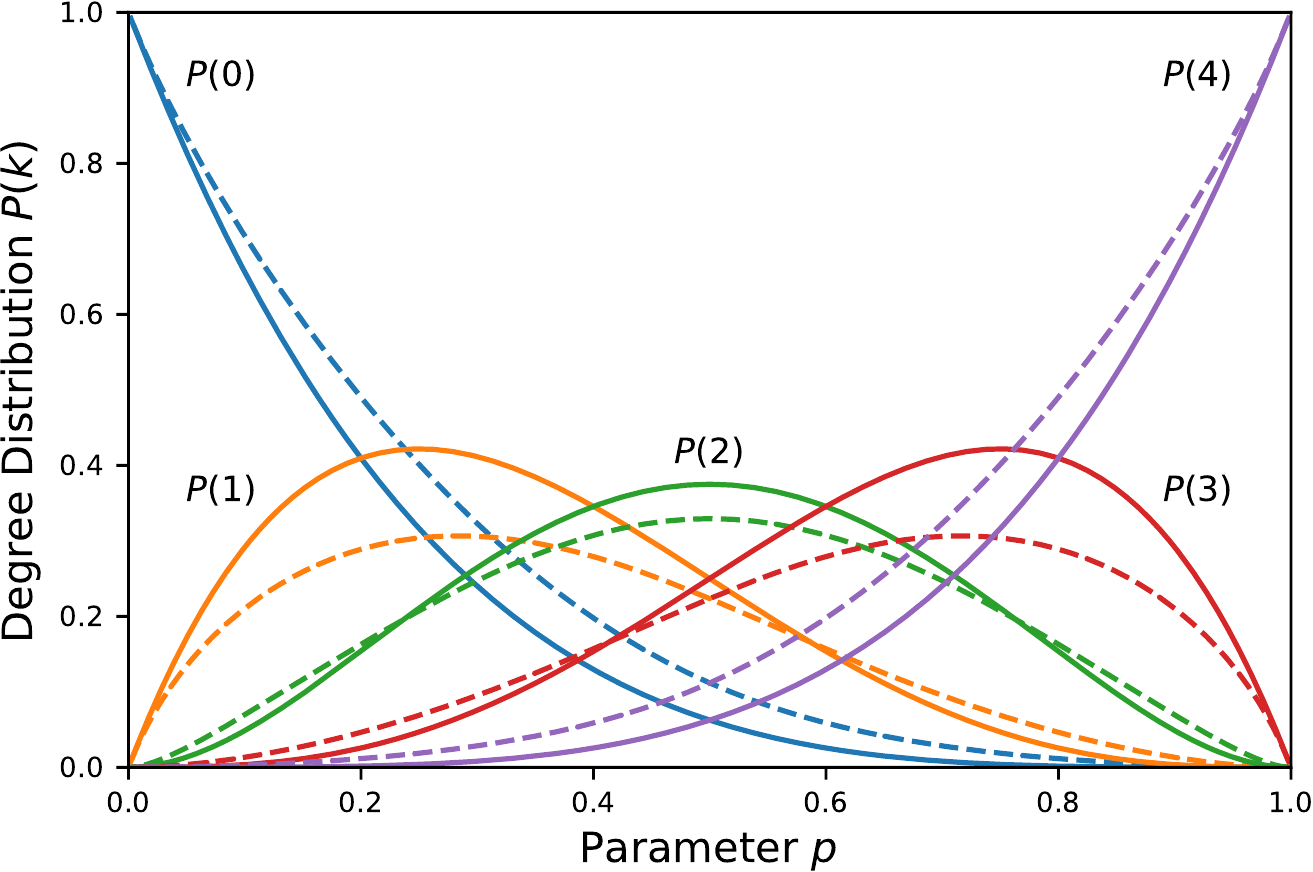}}
	\caption{{\bf The degree distributions in the canonical labeled $\cG_{n, p}$ and unlabeled $\cU_{n, p}$ ER graphs of size $n=5$.} The solid and dashed curves show the exact solutions for the probability $P(k)$ that a random node in a random labeled $\cG_{5, p}$ graph and, respectively, unlabeled $\cU_{5, p}$ graph has degree $k$ for all values of degree $k=0,1,2,3,4$ and density $p \in [0, 1]$.
}
\label{fig:n5}
\end{figure}

\subsubsection{Degree distribution}

Another key difference between labeled/delabeled and unlabeled ER graphs is the degree distribution. As can be deduced from Fig.~\ref{fig:erp}, the degree distribution in the unlabeled graphs is uniform, $P(k)=1/3$ for degrees $k=0,1,2$, versus the corresponding binomial distribution $1/4,1/2,1/4$ in the labeled graphs. Figure~\ref{fig:n5} shows the exact degree distributions in the canonical labeled and unlabeled ER graphs $\cG_{n,p}$ and $\cU_{n,p}$ of size $n=5$. We see that they are different for any values of $p\neq\set{0,1}$. The unlabeled graphs always have more nodes of degree~$0$, for instance.

The degree distribution in $\cU_{n,p}$ is unknown, and we leave it as an open problem to compare it against the degree distribution in~$\cU_{n,m}$, which, as was shown in~\cite{luczak1991deal}, is very different from the Poisson one in sparse $\cG_{n,m}$. In particular, if $m=\bak n/2$ with constant $\bak$, then the $\cU_{n,m}$ graph consists of a connected component of size
\beq
\ell \approx \frac{2m}{\log m},
\eeq
the average degree and degree distribution in which are
\begin{align}
  \bak_{>0} &=\frac{2m}{\ell}\approx\log m\approx\log\ell\approx\log n, \\
  P_\ell(k) & \approx\frac{\log^k\ell}{\ell k!},\quad k>0.
\end{align}
However, most nodes are not in this component and have degree~$0$; their number is $n-\ell\approx n$. The graph is thus dominated by isolated nodes. However, if they are ignored, it has a Poisson-like degree distribution $P_\ell(k)$ with a logarithmically diverging average degree.

\subsubsection{Entropy}

Notwithstanding these drastic structural differences reflected in the degree distribution, the leading terms of entropy of $\cG_{n,m}$ and $\cU_{n,m}$ are surprisingly the same. As can be deduced from~\cite{wright1974graphs}, the entropy of $\cU_{n,m}$ is
\begin{equation}\label{eq:S_U}
  S_U=\frac{\bak}{2}n\log n - \bak n \log\log n + \frac{\bak}{2}(\log\bak - 1)n+o(n)
\end{equation}
for $\bak\ll\log n$. While the leading term is the same,
the subleading terms are different than in the labeled $\cG_{n,m}$ whose entropy for $\bak\ll n$ is~\cite{anand2009entropy}
\begin{equation}\label{eq:S_L}
  S_L=\frac{\bak}{2}n\log n - \frac{\bak}{2}(\log\bak - 1)n+o(\bak n),
\end{equation}
so $S_U<S_L$ for sufficiently large~$n$. Due to the connectivity phase transition at $\bak\sim\log n$, the $\cU_{n,m}$ graphs with $\bak\gg\log n$ do not have any degree-$0$ nodes, and consist of a single connected component, which is asymmetric~\cite{luczak1991deal}.
Since the graph is asymmetric, the labeled and unlabeled entropies are related by $S_U=S_L-\log n!$ in this denser case, but since $\bak\gg\log n$, this difference is negligible.

Nothing is known about the entropy of the canonical $\cU_{n,p}$ model in any regime, including whether there is any ensemble equivalence between $\cU_{n,p}$ and $\cU_{n,m}$ akin to the one established for the labeled $\cG_{n,p}$ and $\cG_{n,m}$~\cite{anand2010gibbs, squartini2015breaking}. We leave these as open problems as well.

\subsection{Configuration model (CM)}

While the entropy of the labeled microcanonical CM is a well explored subject~\cite{bender1978asymptotic, bianconi2008entropy, bianconi2008entropies, bianconi2009entropy, anand2009entropy, wormald2019asymptotic, wegner2021atomic}, very little is known about its unlabeled version. The existing results~\cite{mckay1984automorphisms, brick2020threshold} tell only whether the CM graphs are asymmetric or symmetric, i.e.\ whether they have \emph{any} nontrivial automorphisms, not \emph{how many} automorphisms they have, as is needed for entropy calculations.

As far as sparse power-law degree sequences with exponent $\gamma$ are concerned, the latest results in~\cite{brick2020threshold} show that if $\gamma>3$, then the CM graphs are symmetric. However, it remains unknown what happens for $\gamma\leq3$. Since the key ingredients that break asymmetry (``most graphs are asymmetric''~\cite{cameron2013random}) are hubs in tandem with low-degree nodes (star graphs are ``very symmetric''), the proofs of graph asymmetry (leading to $S_U=S_L-\log n!$) involve strict bounds on the maximum degree and the numbers of nodes of degree $1$ and $2$~\cite{mckay1984automorphisms, brick2020threshold}, which are violated in sparse scale-free degree sequences with $\gamma\leq3$. However, it is still not excluded that such graphs are asymmetric, or, much more likely, that the number of their automorphisms is $\ll n!$. If so, then their unlabeled entropy would be $S_U\approx(\bak/2-1)n\log n$ since $S_L\approx(\bak/2)n\log n$~\cite{bianconi2009entropy}. (Dis)proving this is yet another open problem.\\

Similarly to the unlabeled canonical ER~$\cU_{n,p}$, the unlabeled canonical CM has not been mentioned in the past, so we \emph{define} it here. We first recall that the labeled canonical CM, a.k.a.\ the soft configuration model (SCM)~\cite{park2004statistical, bianconi2008entropy, garlaschelli2008maximum}, is defined by a sequence of expected degrees~$\set{\kappa_i}$ of nodes~$i\in[n]$. The Gibbs probability distribution of random labeled graphs in the model is \beq P(G) = \frac{\exp\left[\sum_{i=1}^n\beta_id_i(G)\right]}{Z}, \eeq where $d_i(G)$ is the degree of node~$i$ in graph~$G\in\cG_n$, and the parameters $\set{\beta_i}$ are found as the solution of the system of $n$ free energy equations \beq \frac{\partial\log Z}{\partial\beta_i}=\sum_{j=1}^{n} p_{ij} = \kappa_i, \eeq where \beq p_{ij} = \frac{1}{e^{-\beta_i-\beta_j}+1} \eeq are the probabilities of edges between nodes~$i$ and~$j$. The model is \emph{not} exchangeable, unless all $\kappa_i$s are the same.

This vanilla SCM definition is clearly not directly applicable to unlabeled graphs since it explicitly refers to node labels~$i$ via~$\kappa_i$. However, the following alternative SCM definition based on the empirical degree distribution in graph~$G\in\cG_n$ avoids this problem: \beq\label{eq:SCM-labeled}  P(G) = \frac{\exp\left[\sum_{k=0}^{n-1}\alpha_kn_k(G)\right]}{Z}.\eeq Here, $n_k(G)$ is the number of nodes of degree~$k$ in~$G$. This version of the labeled SCM \emph{is} exchangeable, and its \emph{definition} is directly applicable to unlabeled graphs $U\in\cU_n$: \beq\label{eq:SCM-unlabeled}  P(U) = \frac{\exp\left[\sum_{k=0}^{n-1}\alpha_kn_k(U)\right]}{Z}.\eeq

As in the ER case, the main difference between the labeled and unlabeled SCMs defined in Eqs.~(\ref{eq:SCM-labeled},\ref{eq:SCM-unlabeled}) is that the partition function~$Z$ involves the summation over all labeled versus unlabeled graphs of size~$n$. Both models are defined by desired expected numbers $\set{\nu_k}$ of nodes of degree~$k$, and the parameters~$\set{\alpha_k}$ are the solutions of the standard free energy equations \beq \frac{\partial\log Z}{\partial\alpha_k}=\nu_k.\eeq

\section{Random geometric graphs (RGGs)}\label{sec:RGG}

Calculating the entropy of RGGs is a cornerstone problem in estimating entropy of a large class of labeled network models with hidden variables~\cite{caldarelli2002scale, soderberg2002general, soderberg2003random, boguna2003class}, where the connection probability between nodes $i$ and $j$ is $p_{ij}=p(x_i,x_j)$, where $p(x,y)\in[0,1]$ is a function of $i,j$'s random coordinates $x_i,x_j$ in some space. These models include not only all spatial networks and latent space models, but also the soft versions of the configuration model, preferential attachment, stochastic block model, as well as graphons in dense graphs. In dense graphs, the number of edges is $m\sim n^2$, so that their entropy, known as graphon entropy~\cite{janson2013graphons}, always dominates the coordinate entropy coming from random $x_i$s. However, in sparse graphs, the numbers of edges and node coordinates are of the same order $\sim n$, so their entropies may be comparable. In RGGs, 
the graph entropy is exactly the coordinate entropy, since given the coordinates, edges exist deterministically.
Therefore, the estimation of the RGG entropy is of utmost importance for disentangling the edge and coordinate entropies.

Here we focus on the sparse one-dimensional RGGs \emph{defined} by
\begin{enumerate}
  \item sprinkling $n$ points uniformly at random over the interval $[0,n]$, and then
  \item linking all pairs of points at distance $<r$ on $[0,n]$, where $r>0$ is a constant,
\end{enumerate}
so the expected average degree converges to $\bak=2r$. Step~(1) implements the binomial point process of rate~$1$ on $[0,n]$, while step~(2) says that $p(x,y)=\indf{|x-y|<r}$, where $\ind$ is the indicator function.

Observe that as defined above, the graphs are actually unlabeled because we did not label the sprinkled points. This is consistent with the general definition of a point process in probability as a random point measure~\cite{kallenberg2017random}, which does not involve any labeling. Denote the entropy of the resulting unlabeled RGGs by~$S_U$.

We can also modify step~($1$) in the definition of unlabeled RGGs above to
\begin{enumerate}
  \item[$1^\prime$.] sample the coordinates $x_i$ of nodes $i\in[n]$ from the uniform distribution on~$[0,n]$ i.i.d.'ly.
\end{enumerate}
The points are now labeled by integers $i\in[n]$, so the resulting graph is labeled as well, but its labels are completely random. That is, it is easy to see that generating the labeled graph using steps~($1^\prime,2$) is equivalent to generating the unlabeled graph using steps~($1,2$) first, and then labeling it by one out of the $n!$ possible permutations of labels~$[n]$ selected uniformly at random. Denote the entropy of the resulting labeled RGGs by~$S_L$.

Going back from the labeled to unlabeled graphs is achieved by generating a labeled graph and then delabeling it. This means that in contrast with ER, the unlabeled and delabeled models of RGGs are actually identical. This is not a surprise but a reflection of the general situation: if a network model is a maximum-entropy null model, as is the case with ER, then its unlabeled and delabeled versions are usually different. However, if a network model is defined by a graph-generation process in which labels do not matter, as is the case with RGGs, then the unlabeled and delabeled versions are identical.

Unfortunately, the entropy of neither unlabeled RGGs~$S_U$ nor labeled ones~$S_L$ is amenable to any brute-force calculations due to the intractable dependencies among edges, so we need to devise some tricks, which are described in the following sections.

\subsection{An upper bound for unlabeled entropy}

First, we upper bound the unlabeled entropy~$S_U$ by the entropy of the uniform distribution over all unlabeled graphs that can be realized as one-dimensional geometric graphs, a.k.a.\ unit interval graphs. This entropy is $\log \cN_U$, where $\cN_U$ is the number of such graphs, which is
\beq
\cN_U=\frac{4^n}{c\sqrt{\pi n^3}},
\eeq
where~$c$ is approximately~$5.01$~\cite{hanlon1982counting}. As a side note, the number of orderly labeled geometric graphs with $x_1<x_2<\ldots<x_n$, whose entropy $S_O$ is squeezed between $S_U$ and $S_L$, $S_U\leq S_O\leq S_L$, is the Catalan number~\cite[Exercise~6.19]{stanley1999enumerative},
\beq
\cN_O=\frac{{2n \choose n}}{n+1}\approx\frac{4^n}{\sqrt{\pi n^3}}.
\eeq
Therefore, our first result is that the entropy of unlabeled RGGs is
\begin{equation}\label{eq:S_U_RGG}
S_U \leq \log\cN_U \approx n\log 4.
\end{equation}

The application of the same technology to $S_L$ would tell us that $S_L \lesssim n \log n$, since $\log\cN_L\sim n\log n$~\cite{hanlon1982counting}, but it would not lead to any lower bound for $S_L$, so it could still be that $S_L\sim n\sim S_U$. We derive much tighter upper and lower bounds for $S_L$ using a different route.

\subsection{An upper bound for labeled entropy}

We first recall a very simple and general relation between the labeled and delabeled entropies~\cite{choi2012compression,kontoyiannis2020compression}. We call the latter the unlabeled entropy below, since unlabeled RGGs are identical to delabeled RGGs.

Consider any model of labeled graphs $G\in\cG_n$ of size~$n$ with distribution $P(G)$ whose entropy is \beq S_L=-\sum_{G\in\cG_n}P(G)\log P(G). \eeq Let $U_G$ be the unlabeled version of $G$, and let $\cG_U$ be the isomorphism class corresponding to an unlabeled graph $U\in\cU_n$: \beq \cG_U=\set{G\in\cG_n:U_G=U}. \eeq Denote its size by $\cN_U=\size{\cG_U}$, and observe that \beq \cN_U=\frac{n!}{\size{\aut(U)}}, \eeq where $\aut(U)$ is the group of automorphisms of any labeled version of~$U$. Let $P(U)$ be the delabeled probability distribution induced by $P(G)$, \beq P(U)=\sum_{G\in\cG_U}P(G). \eeq Observe that since any labeled graph~$G$ has only one unlabeled graph~$U_G$ corresponding to it, we have that \beq P(G) = \sum_{U \in \mathcal{U}_n}P(G|U)P(U) = P(G|U_G)P(U_G). \eeq From here it follows that the conditional distribution of graph~$G\in\cG_U$ given that its unlabeled graph is~$U_G$ is \beq P(G|U_G)=\frac{P(G)}{P(U_G)}. \eeq  If the model is exchangeable, as is the case with labeled RGGs, then $P(G|U_G)$ is uniform, \beq P(G|U_G)=\frac{1}{\cN_{U_G}}, \eeq hence \beq P(G)=\frac{P(U_G)}{\cN_{U_G}}. \eeq Substituting this into $S_L$ yields
\begin{align}
  S_L &= -\sum_{G\in\cG_n}P(G)\log\left[\frac{P(U_G)}{\cN_{U_G}}\right] \nonumber\\
      &=\log n!-\sum_{U\in\cU_n}\sum_{G\in\cG_U}P(G)[\log P(U)+\log\size{\aut(U)}] \nonumber\\
      &=\log n!-\sum_{U\in\cU_n}P(U)\log P(U) \nonumber \\
      &-\sum_{U\in\cU_n}P(U)\log\size{\aut(U)}.
\end{align}
That is,
\begin{equation}\label{SL.vs.SU}
  S_L=S_U+\log n!-\cA,
\end{equation}
where \beq S_U=-\sum_{U\in\cU_n}P(U)\log P(U) \eeq is the unlabeled entropy, and \beq \cA=\ea{\log\size{\aut(U)}}=\sum_{U\in\cU_n}P(U)\log\size{\aut(U)} \eeq is the expected logsize of the automorphism group.

Equation~\eqref{SL.vs.SU} provides the following useful upper and lower bounds for the labeled entropy:
\begin{align}
  S_L &\leq S_U + \log n!, \label{eq:S_L_upper}\\
  S_L &\geq \log n! - \cA. \label{eq:S_L_lower}
\end{align}
Since $S_U\lesssim n$ in our RGGs, we immediately arrive at the upper bound for their labeled entropy using~\eqref{eq:S_L_upper}: 
\beq S_L\leq n \log n. \eeq

An upper bound on~$\cA$ would yield a lower bound on~$S_L$ using~\eqref{eq:S_L_lower}, which we deal with next.

\subsection{A lower bound for labeled entropy}

We first assume that $n$ is sufficiently large, so we can approximate the binomial point process of rate~$1$ on $[0,n]$ with the Poisson one, where the distances~$d$ between consecutive points are independent exponentially distributed random variables with PDF $P(d)=e^{-d}$~\cite{last2017lectures}. We then recall that the percolation threshold in one-dimensional RGGs is infinite, simply because $d>r$ with probability $p=e^{-r}$. It follows that the sizes $s_c$ of connected components $c\in[C]$ are independent geometrically distributed random variables with PDF $P(s)=p(1-p)^{s-1}$, while the number of components $C$ is approximately binomial, $P(C)={n-1\choose C-1}p^{C-1}(1-p)^{n-C}$, $C\in[n]$.

The key observation then is that for a label permutation to be an automorphism, it must either permute nodes within a component, or permute the components, or both. A trivial upper bound \beq A=C!\prod_{c=1}^C s_c! \eeq for the number of automorphisms is when all components~$c$ are maximally symmetric, i.e.\ when they all are complete graphs of size~$s_c$. It follows that \beq \log A \approx C\log C - C + \sum_{c=1}^{C}\log s_c!, \eeq and since $\ea{C}\approx pn$ and $\ea{s_c}=1/p=e^r$, we see that the leading term in $\ea{\log A}$ is $\ea{C\log C}$, which one can check is $\approx pn\log n$. 
We thus have that \beq \cA\leq pn\log n. \eeq Substituting this into~\eqref{eq:S_L_lower}, and combining with the upper bound obtained earlier, we finally get
\begin{equation}\label{eq:S_L_RGG}
  (1-e^{-\bak/2}) n \log n \leq S_L \leq n \log n.
\end{equation}

We note that the larger the average degree, the tighter these bounds, although they are asymptotic, holding for $\ea{C}\gg1$, meaning $n \gg e^{\bak/2}$.

Comparing~\eqref{eq:S_L_RGG} with~\eqref{eq:S_U_RGG}, we conclude that $S_U \ll S_L$. That is, the entropy of random labeling $\log n!$ in~(\ref{eq:S_L_upper},\ref{eq:S_L_lower}) dominates the network-structural entropy~\eqref{eq:S_U_RGG}.

\section{Conclusions}\label{sec:conclusion}

Null models of networks are typically used to assess the statistical significance of network features in a given real-world network, and to investigate whether such features are relevant for a particular set of network functions~\cite{orsini2015quantifying}. Colloquially, null models are ``maximally random'' models of networks constrained to have a set of particular network properties. Real networks are compared against their null models to detect statistically significant deviations in the values of network properties that are not constrained in the null model. The ``maximum randomness'' of null models is formally achieved by maximizing the model entropy under the imposed constraints. But should these models be models of labeled or unlabeled networks? This elephant-in-the-room question has been magically ignored in the past research.

Here we began with a crucial observation that if nodes in a real-world network are labeled and \emph{distinguishable}, then correct null models for such a network must be either \emph{nonexchangeable or unlabeled}, simply because exchangeability is statistical indistinguishability. Exchangeable models represent a world which is statistically very different from any data source generating distinguishable labels. And if we are interested only in the network structure, then the correct null models of the structure of real networks with distinguishable nodes must be models of unlabeled graphs.

The logical next two questions are then:
\begin{enumerate}
 \item Does it really matter what models to consider, labeled or unlabeled, as they may be equivalent upon some simple transformation?
 \item If it does matter, and unlabeled models should have been studied all these years as much as---if not more than---labeled ones, then why has this not been the case?
\end{enumerate}

To address the first question, we have shown here that labeled and unlabeled models of sparse networks can be very different and nonequivalent. The emphasis on \emph{sparse} networks is important here. Almost all \emph{dense} networks are asymmetric~\cite{cameron2013random}. Therefore, there exists a simple relation between their labeled and unlabeled models: almost every unlabeled network corresponds to exactly $n!$ labeled ones with all possible permutations of labels. It follows then that as far as entropy is concerned, for instance, the general equation~\eqref{SL.vs.SU} relating the labeled and unlabeled network entropies $S_L$ and $S_U$ becomes trivial: $S_L \approx S_U + \log n!$. But since the unlabeled entropy of dense networks is $S_U \sim n^2$~\cite{janson2013graphons, hoorn2018sparse}, this difference between $S_L$ and $S_U$ is negligible, $S_L\sim S_U \sim n^2$. In other words, in dense networks, there is room for huge diversity of network structure. Therefore, randomness associated with network structure, represented by the unlabeled entropy~$S_U\sim n^2$, dominates the labeling entropy~$S_\cL=S_L-S_U\approx n\log n$. 

We have shown that in sparse networks, the situation is very different. The sources of network-structural and labeling randomnesses are comparable in their power, so the network-structural entropy~$S_U$ may or may not be the leading factor, depending on the model. We have considered three examples demonstrating all of the three possibilities:
\begin{itemize}
  \item in \erdren graphs, the network-structural entropy dominates the labeling entropy:\\$S_U\gg S_\cL$;
  \item in the scale-free configuration model, the two entropies are conjecturally comparable:\\$S_U \sim S_\cL$; 
  \item in random geometric graphs, the labeling entropy dominates the network-structural entropy:\\$S_U \ll S_\cL$.
\end{itemize}

Even though the network-structural entropy wins in sparse \erdren graphs, this example is still a source of concerns for at least two reasons. First, by definition, \erdren graphs are maximally random graphs with a given average degree, so this is where we could expect to see the strongest domination of~$S_U$ over~$S_\cL$, compared to other sparse network models. However, this domination is not really strong. In fact, it is marginal: as can be seen from~(\ref{eq:S_U},\ref{eq:S_L}), $S_U/S_\cL \approx (1/2)\log n/\log\log n$.

Worse, the entropic equivalence is definitely a necessary but not sufficient condition for model equivalence. The \erdren example shows that even though the labeled and unlabeled models are entropically equivalent ($S_L\approx S_U$), their very basic structural property---the degree distribution---is very different between the two models.

Compared to \erdren graphs, the random geometric graph example is much more disconcerting as it clearly demonstrates that even the entropic equivalence between labeled and unlabeled models can be broken in sparse networks, and that the entropy of meaningless labeling noise~$S_\cL$ may easily be the leading factor, overpowering the network-structural signal~$S_U$. This result is disconcerting because it implies that entropy maximization in sparse networks may easily be the maximization of the entropy of meaningless labeling that you do not care about, versus the intended maximization of the entropy of the network structure. Such caveats can easily lead to profound aberrations and statistical errors in conclusions made about the structure of sparse real-world networks based on their maximum-entropy labeled models.

It is important to note that in some situations, working with labeled network models is justified. As mentioned in Section~\ref{sec:background}, models of real-world networks with indistinguishable nodes must be labeled. Other examples include situations where labeling is statistically meaningful. One such example, briefly mentioned in Section~\ref{sec:RGG}, is the random geometric graph labeled in the order of increasing coordinates. Thus labeled, random geometric graphs can be formulated as a growing network model~\cite{krioukov2013duality}. This example generalizes to any growing network model, such as preferential attachment whose symmetry properties were considered in~\cite{luczak2019asymmetry}, where the preferred labeling is by nodes' birth times. Yet another example of a different sort is the stochastic block model where nodes can be labeled by communities they belong to~\cite{peixoto2012entropy, peixoto2013parsimonious}.

Unfortunately, even in cases where there exists a (unique) preferred labeling scheme, the models that have been actually studied at depth in the past are the models of networks labeled by~$[n]$ arbitrarily. Even more unfortunate is that this practice is nearly never spelt out explicitly. As a rule of thumb, if unsure what networks, labeled or unlabeled, a particular result in network science or graph theory is about, assume it is about arbitrarily labeled networks.

Perhaps the \emph{main implication} of the results in this paper is that this practice of silence about the differences between labeled and unlabeled network models should be abandoned, at least as far as applications to sparse networks are concerned. It is really imperative to understand in what situations the replacement of the correct representation of the network structure, an unlabeled network, by its simpler labeled surrogate, is statistically justified. As a bare minimum, it should always be made clear what networks we are dealing with, labeled or unlabeled, and why. As we have seen here, the answer to this \emph{why} question may be very difficult. It may not even exist, suggesting a dire need for a thorough reexamination of the foundational results in the science of sparse networks. To motivate this reexamination, note that its very basic starting point, the degree distribution in unlabeled \erdren graphs, is not what you would have expected.\\

Given that the very core of network science is all about the \emph{structure} (and function) of complex networks~\cite{newman2003structure}, one may ask why all of the best-studied null models of the structure of networks are labeled, not unlabeled as they should have been. Why did unlabeled graphs not attract the deserved attention even in textbooks on network science and graph theory? We speculate that the main reason is that it is actually quite difficult not only to think about unlabeled networks, but also to deal with them in practice.

First of all, it is quite a challenge to store an unlabeled graph on a computer. The standard practice to represent an unlabeled graph in textbooks, web sites, or computer programs like \texttt{Mathematica}, is to picture it. Yet storing graphs in pictures does not get you far computationally, so \texttt{Mathematica} and similar programs rely on \emph{canonical labeling}~\cite{mckay2014practical} of unlabeled graphs, an important area of research in graph theory and computer science looking for computationally efficient ways to assign a unique labeling to an unlabeled graph.

This ``mundane'' graph representation issue is a nuance, compared to the problem of graph generation. It is quite a challenge to generate even the simplest unlabeled graphs, i.e.\ unlabeled \erdren graphs. Some of the best results on how to generate microcanonical unlabeled \erdren graphs are available in~\cite{wormald1987generating}. Since the canonical unlabeled \erdren graphs have not been considered or even mentioned before, there are no results whatsoever on how to generate them. Nor are we aware of any results on how to generate graphs in the unlabeled configuration model, either microcanonical or canonical, the latter also defined for the first time in this paper only.

Yet it is not the case that there are no strongly positive results on unlabeled graphs at all. The optimistic results in~\cite{luczak1991deal} say that in microcanonical unlabeled \erdren graphs, the values of a huge class of network properties can be linked to their values in the corresponding labeled graphs. Roughly, the unlabeled values are the labeled values in graphs of a different effective size. Unfortunately, these results apply mostly to denser graphs with the average degree $\bak \gg \log n$.

Another source of optimism is that sufficient statistics in exchangeable models are unlabeled properties. By {\it unlabeled properties}, we mean those network properties whose values are identical across all labeled graphs in any isomorphism class. Therefore, there exists a unique unlabeled model corresponding to each exchangeable model, and vice versa, having identical sufficient statistics, differing only in their sample spaces, i.e., unlabeled vs.\ labeled graphs. As discussed in Section~\ref{sec:background}, if all of the network properties that a researcher has data for and wants to model in a null model are unlabeled, then a natural setting for the sample space would be that of unlabeled graphs. Regardless of the choice of an unlabeled versus exchangeably labeled model in such cases, the discussed connection between these two types of models may help to translate ideas, methods, and techniques between the two settings.

To end on a truly positive note, the last remark is that unlabeled network models completely avoid the exchangeability conundrum in sparse networks. On the one hand, if node labels are meaningless integers that ``do not matter,'' then the network model must be exchangeable since the probability of a network in the model cannot depend on how the network is labeled. On the other hand, the thermodynamic limit of any sparse exchangeable network model necessarily consists of empty networks due to the Aldous-Hoover theorem~\cite{hoover1979relations, aldous1981representations}. Exchangeability makes no sense in the realm of unlabeled graphs, so the paradox dissolves there. The Aldous-Hoover theorem simply implies that the limits of sparse unlabeled networks---whatever they are---cannot be exchangeably labeled by integers.

\begin{acknowledgments}
We thank O.~Kallenberg, Y.~Peres, M.~R\'acz, Z.~Burda, P.~Krapivsky, N.~Wormald, T.~{\L}uczak, F.~Radicchi, and G.~Bianconi for useful discussions and suggestions. This work was supported by the NSF grant IIS-1741355.
\end{acknowledgments}

\onecolumngrid

\end{document}

%% file: defs.tex
\usepackage[utf8]{inputenc}
\usepackage{amsfonts,amsmath,amssymb,amsthm,mathtools,bbm,graphicx}
\usepackage[svgnames,dvipsnames,x11names]{xcolor}
\usepackage[colorlinks=true,allcolors=blue]{hyperref}

\renewcommand{\leq}{\leqslant}
\renewcommand{\geq}{\geqslant}

\DeclareMathOperator{\aut}{Aut}

\def\R{\mathbb{R}}

\def\cA{\mathcal{A}}

\def\cD{\mathcal{D}}

\def\cG{\mathcal{G}}

\def\cL{\mathcal{L}}

\def\cN{\mathcal{N}}

\def\cU{\mathcal{U}}

\def\cL{\mathcal{L}}

\newcommand{\ind}{\boldsymbol{\mathbbm{1}}} 
\newcommand{\indf}[1]{\ind\set{#1}} 

\newcommand{\size}[1]{\left|#1\right|}

\newcommand{\set}[1]{\left\{#1\right\}}

\providecommand{\setthms}[1]{#1}
\setthms{

\theoremstyle{definition}

}

\newcommand{\erdren}{Erd\H{o}s-R\'enyi }

\definecolor{mygreen}{rgb}{0, 0.68, 0.31}
\definecolor{myred}{rgb}{1.0, 0,0}

\def\bsplit#1\esplit{\begin{split} #1 \end{split} }
\def\splitb#1\splite{\begin{split} #1 \end{split} }
\def\beq#1\eeq{\begin{equation} #1 \end{equation}}
\def\eqb#1\eqe{\begin{equation} #1 \end{equation}}

\newcommand{\ea}[1]{\left\langle{#1}\right\rangle}

\renewcommand{\epsilon}{\varepsilon}

\def\bak{\bar{k}}
\def\bam{\bar{m}}
\def\bad{\bar{d}}

\def\denote{\coloneqq} 

%% file: paper.bbl
\begin{thebibliography}{65}%
\makeatletter
\providecommand \@ifxundefined [1]{%
 \@ifx{#1\undefined}
}%
\providecommand \@ifnum [1]{%
 \ifnum #1\expandafter \@firstoftwo
 \else \expandafter \@secondoftwo
 \fi
}%
\providecommand \@ifx [1]{%
 \ifx #1\expandafter \@firstoftwo
 \else \expandafter \@secondoftwo
 \fi
}%
\providecommand \natexlab [1]{#1}%
\providecommand \enquote  [1]{``#1''}%
\providecommand \bibnamefont  [1]{#1}%
\providecommand \bibfnamefont [1]{#1}%
\providecommand \citenamefont [1]{#1}%
\providecommand \href@noop [0]{\@secondoftwo}%
\providecommand \href [0]{\begingroup \@sanitize@url \@href}%
\providecommand \@href[1]{\@@startlink{#1}\@@href}%
\providecommand \@@href[1]{\endgroup#1\@@endlink}%
\providecommand \@sanitize@url [0]{\catcode `\\12\catcode `\$12\catcode
  `\&12\catcode `\#12\catcode `\^12\catcode `\_12\catcode `\%12\relax}%
\providecommand \@@startlink[1]{}%
\providecommand \@@endlink[0]{}%
\providecommand \url  [0]{\begingroup\@sanitize@url \@url }%
\providecommand \@url [1]{\endgroup\@href {#1}{\urlprefix }}%
\providecommand \urlprefix  [0]{URL }%
\providecommand \Eprint [0]{\href }%
\providecommand \doibase [0]{http://dx.doi.org/}%
\providecommand \selectlanguage [0]{\@gobble}%
\providecommand \bibinfo  [0]{\@secondoftwo}%
\providecommand \bibfield  [0]{\@secondoftwo}%
\providecommand \translation [1]{[#1]}%
\providecommand \BibitemOpen [0]{}%
\providecommand \bibitemStop [0]{}%
\providecommand \bibitemNoStop [0]{.\EOS\space}%
\providecommand \EOS [0]{\spacefactor3000\relax}%
\providecommand \BibitemShut  [1]{\csname bibitem#1\endcsname}%
\let\auto@bib@innerbib\@empty
\bibitem [{\citenamefont {Park}\ and\ \citenamefont
  {Newman}(2004)}]{park2004statistical}%
  \BibitemOpen
  \bibfield  {author} {\bibinfo {author} {\bibfnamefont {J.}~\bibnamefont
  {Park}}\ and\ \bibinfo {author} {\bibfnamefont {M.~E.~J.}\ \bibnamefont
  {Newman}},\ }\bibfield  {title} {\emph {\bibinfo {title} {{Statistical
  mechanics of networks}},\ }}\href {\doibase 10.1103/PhysRevE.70.066117}
  {\bibfield  {journal} {\bibinfo  {journal} {Phys Rev E}\ }\textbf {\bibinfo
  {volume} {70}},\ \bibinfo {pages} {066117} (\bibinfo {year}
  {2004})}\BibitemShut {NoStop}%
\bibitem [{\citenamefont {Bianconi}(2008)}]{bianconi2008entropy}%
  \BibitemOpen
  \bibfield  {author} {\bibinfo {author} {\bibfnamefont {G.}~\bibnamefont
  {Bianconi}},\ }\bibfield  {title} {\emph {\bibinfo {title} {{The entropy of
  randomized network ensembles}},\ }}\href {\doibase
  10.1209/0295-5075/81/28005} {\bibfield  {journal} {\bibinfo  {journal} {EPL}\
  }\textbf {\bibinfo {volume} {81}},\ \bibinfo {pages} {28005} (\bibinfo {year}
  {2008})}\BibitemShut {NoStop}%
\bibitem [{\citenamefont {Bianconi}\ \emph {et~al.}(2008)\citenamefont
  {Bianconi}, \citenamefont {Coolen},\ and\ \citenamefont {{Perez
  Vicente}}}]{bianconi2008entropies}%
  \BibitemOpen
  \bibfield  {author} {\bibinfo {author} {\bibfnamefont {G.}~\bibnamefont
  {Bianconi}}, \bibinfo {author} {\bibfnamefont {A.~C.~C.}\ \bibnamefont
  {Coolen}}, \ and\ \bibinfo {author} {\bibfnamefont {C.}~\bibnamefont {{Perez
  Vicente}}},\ }\bibfield  {title} {\emph {\bibinfo {title} {{Entropies of
  complex networks with hierarchically constrained topologies}},\ }}\href
  {\doibase 10.1103/PhysRevE.78.016114} {\bibfield  {journal} {\bibinfo
  {journal} {Phys Rev E}\ }\textbf {\bibinfo {volume} {78}},\ \bibinfo {pages}
  {016114} (\bibinfo {year} {2008})}\BibitemShut {NoStop}%
\bibitem [{\citenamefont {Bianconi}(2009)}]{bianconi2009entropy}%
  \BibitemOpen
  \bibfield  {author} {\bibinfo {author} {\bibfnamefont {G.}~\bibnamefont
  {Bianconi}},\ }\bibfield  {title} {\emph {\bibinfo {title} {{Entropy of
  network ensembles}},\ }}\href {\doibase 10.1103/PhysRevE.79.036114}
  {\bibfield  {journal} {\bibinfo  {journal} {Phys Rev E}\ }\textbf {\bibinfo
  {volume} {79}},\ \bibinfo {pages} {036114} (\bibinfo {year}
  {2009})}\BibitemShut {NoStop}%
\bibitem [{\citenamefont {Anand}\ and\ \citenamefont
  {Bianconi}(2009)}]{anand2009entropy}%
  \BibitemOpen
  \bibfield  {author} {\bibinfo {author} {\bibfnamefont {K.}~\bibnamefont
  {Anand}}\ and\ \bibinfo {author} {\bibfnamefont {G.}~\bibnamefont
  {Bianconi}},\ }\bibfield  {title} {\emph {\bibinfo {title} {{Entropy measures
  for networks: Toward an information theory of complex topologies}},\ }}\href
  {\doibase 10.1103/PhysRevE.80.045102} {\bibfield  {journal} {\bibinfo
  {journal} {Phys Rev E}\ }\textbf {\bibinfo {volume} {80}},\ \bibinfo {pages}
  {045102} (\bibinfo {year} {2009})}\BibitemShut {NoStop}%
\bibitem [{\citenamefont {Anand}\ \emph {et~al.}(2011)\citenamefont {Anand},
  \citenamefont {Bianconi},\ and\ \citenamefont {Severini}}]{anand2011shannon}%
  \BibitemOpen
  \bibfield  {author} {\bibinfo {author} {\bibfnamefont {K.}~\bibnamefont
  {Anand}}, \bibinfo {author} {\bibfnamefont {G.}~\bibnamefont {Bianconi}}, \
  and\ \bibinfo {author} {\bibfnamefont {S.}~\bibnamefont {Severini}},\
  }\bibfield  {title} {\emph {\bibinfo {title} {{Shannon and von Neumann
  entropy of random networks with heterogeneous expected degree}},\ }}\href
  {\doibase 10.1103/PhysRevE.83.036109} {\bibfield  {journal} {\bibinfo
  {journal} {Phys Rev E}\ }\textbf {\bibinfo {volume} {83}},\ \bibinfo {pages}
  {036109} (\bibinfo {year} {2011})}\BibitemShut {NoStop}%
\bibitem [{\citenamefont {Garlaschelli}\ and\ \citenamefont
  {Loffredo}(2008)}]{garlaschelli2008maximum}%
  \BibitemOpen
  \bibfield  {author} {\bibinfo {author} {\bibfnamefont {D.}~\bibnamefont
  {Garlaschelli}}\ and\ \bibinfo {author} {\bibfnamefont {M.}~\bibnamefont
  {Loffredo}},\ }\bibfield  {title} {\emph {\bibinfo {title} {{Maximum
  likelihood: Extracting unbiased information from complex networks}},\ }}\href
  {\doibase 10.1103/PhysRevE.78.015101} {\bibfield  {journal} {\bibinfo
  {journal} {Phys Rev E}\ }\textbf {\bibinfo {volume} {78}},\ \bibinfo {pages}
  {015101} (\bibinfo {year} {2008})}\BibitemShut {NoStop}%
\bibitem [{\citenamefont {Garlaschelli}\ and\ \citenamefont
  {Loffredo}(2009)}]{garlaschelli2009generalized}%
  \BibitemOpen
  \bibfield  {author} {\bibinfo {author} {\bibfnamefont {D.}~\bibnamefont
  {Garlaschelli}}\ and\ \bibinfo {author} {\bibfnamefont {M.}~\bibnamefont
  {Loffredo}},\ }\bibfield  {title} {\emph {\bibinfo {title} {{Generalized
  Bose-Fermi Statistics and Structural Correlations in Weighted Networks}},\
  }}\href {\doibase 10.1103/PhysRevLett.102.038701} {\bibfield  {journal}
  {\bibinfo  {journal} {Phys Rev Lett}\ }\textbf {\bibinfo {volume} {102}},\
  \bibinfo {pages} {038701} (\bibinfo {year} {2009})}\BibitemShut {NoStop}%
\bibitem [{\citenamefont {Squartini}\ and\ \citenamefont
  {Garlaschelli}(2011)}]{squartini2011analytical}%
  \BibitemOpen
  \bibfield  {author} {\bibinfo {author} {\bibfnamefont {T.}~\bibnamefont
  {Squartini}}\ and\ \bibinfo {author} {\bibfnamefont {D.}~\bibnamefont
  {Garlaschelli}},\ }\bibfield  {title} {\emph {\bibinfo {title} {{Analytical
  maximum-likelihood method to detect patterns in real networks}},\ }}\href
  {\doibase 10.1088/1367-2630/13/8/083001} {\bibfield  {journal} {\bibinfo
  {journal} {New J Phys}\ }\textbf {\bibinfo {volume} {13}},\ \bibinfo {pages}
  {083001} (\bibinfo {year} {2011})}\BibitemShut {NoStop}%
\bibitem [{\citenamefont {Mastrandrea}\ \emph {et~al.}(2014)\citenamefont
  {Mastrandrea}, \citenamefont {Squartini}, \citenamefont {Fagiolo},\ and\
  \citenamefont {Garlaschelli}}]{mastrandrea2014enhanced}%
  \BibitemOpen
  \bibfield  {author} {\bibinfo {author} {\bibfnamefont {R.}~\bibnamefont
  {Mastrandrea}}, \bibinfo {author} {\bibfnamefont {T.}~\bibnamefont
  {Squartini}}, \bibinfo {author} {\bibfnamefont {G.}~\bibnamefont {Fagiolo}},
  \ and\ \bibinfo {author} {\bibfnamefont {D.}~\bibnamefont {Garlaschelli}},\
  }\bibfield  {title} {\emph {\bibinfo {title} {{Enhanced reconstruction of
  weighted networks from strengths and degrees}},\ }}\href {\doibase
  10.1088/1367-2630/16/4/043022} {\bibfield  {journal} {\bibinfo  {journal}
  {New J Phys}\ }\textbf {\bibinfo {volume} {16}},\ \bibinfo {pages} {043022}
  (\bibinfo {year} {2014})}\BibitemShut {NoStop}%
\bibitem [{\citenamefont {Squartini}\ \emph
  {et~al.}(2015{\natexlab{a}})\citenamefont {Squartini}, \citenamefont
  {Mastrandrea},\ and\ \citenamefont {Garlaschelli}}]{squartini2015unbiased}%
  \BibitemOpen
  \bibfield  {author} {\bibinfo {author} {\bibfnamefont {T.}~\bibnamefont
  {Squartini}}, \bibinfo {author} {\bibfnamefont {R.}~\bibnamefont
  {Mastrandrea}}, \ and\ \bibinfo {author} {\bibfnamefont {D.}~\bibnamefont
  {Garlaschelli}},\ }\bibfield  {title} {\emph {\bibinfo {title} {{Unbiased
  sampling of network ensembles}},\ }}\href {\doibase
  10.1088/1367-2630/17/2/023052} {\bibfield  {journal} {\bibinfo  {journal}
  {New J Phys}\ }\textbf {\bibinfo {volume} {17}},\ \bibinfo {pages} {023052}
  (\bibinfo {year} {2015}{\natexlab{a}})}\BibitemShut {NoStop}%
\bibitem [{\citenamefont {Zuev}\ \emph {et~al.}(2015)\citenamefont {Zuev},
  \citenamefont {Eisenberg},\ and\ \citenamefont
  {Krioukov}}]{zuev2015exponential}%
  \BibitemOpen
  \bibfield  {author} {\bibinfo {author} {\bibfnamefont {K.}~\bibnamefont
  {Zuev}}, \bibinfo {author} {\bibfnamefont {O.}~\bibnamefont {Eisenberg}}, \
  and\ \bibinfo {author} {\bibfnamefont {D.}~\bibnamefont {Krioukov}},\
  }\bibfield  {title} {\emph {\bibinfo {title} {{Exponential Random Simplicial
  Complexes}},\ }}\href {\doibase 10.1088/1751-8113/48/46/465002} {\bibfield
  {journal} {\bibinfo  {journal} {J Phys A Math Theor}\ }\textbf {\bibinfo
  {volume} {48}},\ \bibinfo {pages} {21} (\bibinfo {year} {2015})}\BibitemShut
  {NoStop}%
\bibitem [{\citenamefont {van~der Hoorn}\ \emph {et~al.}(2018)\citenamefont
  {van~der Hoorn}, \citenamefont {Lippner},\ and\ \citenamefont
  {Krioukov}}]{hoorn2018sparse}%
  \BibitemOpen
  \bibfield  {author} {\bibinfo {author} {\bibfnamefont {P.}~\bibnamefont
  {van~der Hoorn}}, \bibinfo {author} {\bibfnamefont {G.}~\bibnamefont
  {Lippner}}, \ and\ \bibinfo {author} {\bibfnamefont {D.}~\bibnamefont
  {Krioukov}},\ }\bibfield  {title} {\emph {\bibinfo {title} {{Sparse
  Maximum-Entropy Random Graphs with a Given Power-Law Degree Distribution}},\
  }}\href {\doibase 10.1007/s10955-017-1887-7} {\bibfield  {journal} {\bibinfo
  {journal} {J Stat Phys}\ }\textbf {\bibinfo {volume} {173}},\ \bibinfo
  {pages} {806} (\bibinfo {year} {2018})}\BibitemShut {NoStop}%
\bibitem [{\citenamefont {Anand}\ and\ \citenamefont
  {Bianconi}(2010)}]{anand2010gibbs}%
  \BibitemOpen
  \bibfield  {author} {\bibinfo {author} {\bibfnamefont {K.}~\bibnamefont
  {Anand}}\ and\ \bibinfo {author} {\bibfnamefont {G.}~\bibnamefont
  {Bianconi}},\ }\bibfield  {title} {\emph {\bibinfo {title} {{Gibbs entropy of
  network ensembles by cavity methods}},\ }}\href {\doibase
  10.1103/PhysRevE.82.011116} {\bibfield  {journal} {\bibinfo  {journal} {Phys
  Rev E}\ }\textbf {\bibinfo {volume} {82}},\ \bibinfo {pages} {011116}
  (\bibinfo {year} {2010})}\BibitemShut {NoStop}%
\bibitem [{\citenamefont {Squartini}\ \emph
  {et~al.}(2015{\natexlab{b}})\citenamefont {Squartini}, \citenamefont
  {de~Mol}, \citenamefont {den Hollander},\ and\ \citenamefont
  {Garlaschelli}}]{squartini2015breaking}%
  \BibitemOpen
  \bibfield  {author} {\bibinfo {author} {\bibfnamefont {T.}~\bibnamefont
  {Squartini}}, \bibinfo {author} {\bibfnamefont {J.}~\bibnamefont {de~Mol}},
  \bibinfo {author} {\bibfnamefont {F.}~\bibnamefont {den Hollander}}, \ and\
  \bibinfo {author} {\bibfnamefont {D.}~\bibnamefont {Garlaschelli}},\
  }\bibfield  {title} {\emph {\bibinfo {title} {{Breaking of Ensemble
  Equivalence in Networks}},\ }}\href {\doibase 10.1103/PhysRevLett.115.268701}
  {\bibfield  {journal} {\bibinfo  {journal} {Phys Rev Lett}\ }\textbf
  {\bibinfo {volume} {115}},\ \bibinfo {pages} {268701} (\bibinfo {year}
  {2015}{\natexlab{b}})}\BibitemShut {NoStop}%
\bibitem [{\citenamefont {Garlaschelli}\ \emph {et~al.}(2017)\citenamefont
  {Garlaschelli}, \citenamefont {den Hollander},\ and\ \citenamefont
  {Roccaverde}}]{garlaschelli2017ensemble}%
  \BibitemOpen
  \bibfield  {author} {\bibinfo {author} {\bibfnamefont {D.}~\bibnamefont
  {Garlaschelli}}, \bibinfo {author} {\bibfnamefont {F.}~\bibnamefont {den
  Hollander}}, \ and\ \bibinfo {author} {\bibfnamefont {A.}~\bibnamefont
  {Roccaverde}},\ }\bibfield  {title} {\emph {\bibinfo {title} {{Ensemble
  nonequivalence in random graphs with modular structure}},\ }}\href {\doibase
  10.1088/1751-8113/50/1/015001} {\bibfield  {journal} {\bibinfo  {journal} {J
  Phys A Math Theor}\ }\textbf {\bibinfo {volume} {50}},\ \bibinfo {pages}
  {015001} (\bibinfo {year} {2017})}\BibitemShut {NoStop}%
\bibitem [{\citenamefont {Shannon}(1948)}]{shannon1948mathematical}%
  \BibitemOpen
  \bibfield  {author} {\bibinfo {author} {\bibfnamefont {C.~E.}\ \bibnamefont
  {Shannon}},\ }\bibfield  {title} {\emph {\bibinfo {title} {{A Mathematical
  Theory of Communication}},\ }}\href {\doibase
  10.1002/j.1538-7305.1948.tb01338.x} {\bibfield  {journal} {\bibinfo
  {journal} {Bell Syst Tech J}\ }\textbf {\bibinfo {volume} {27}},\ \bibinfo
  {pages} {379} (\bibinfo {year} {1948})}\BibitemShut {NoStop}%
\bibitem [{\citenamefont {Cover}\ and\ \citenamefont
  {Thomas}(2005)}]{cover2005elements}%
  \BibitemOpen
  \bibfield  {author} {\bibinfo {author} {\bibfnamefont {T.~M.}\ \bibnamefont
  {Cover}}\ and\ \bibinfo {author} {\bibfnamefont {J.~A.}\ \bibnamefont
  {Thomas}},\ }\href {\doibase 10.1002/047174882X} {\emph {\bibinfo {title}
  {{Elements of Information Theory}}}}\ (\bibinfo  {publisher} {Wiley},\
  \bibinfo {address} {Hoboken, NJ},\ \bibinfo {year} {2005})\BibitemShut
  {NoStop}%
\bibitem [{\citenamefont {Bianconi}\ \emph {et~al.}(2009)\citenamefont
  {Bianconi}, \citenamefont {Pin},\ and\ \citenamefont
  {Marsili}}]{bianconi2009assessing}%
  \BibitemOpen
  \bibfield  {author} {\bibinfo {author} {\bibfnamefont {G.}~\bibnamefont
  {Bianconi}}, \bibinfo {author} {\bibfnamefont {P.}~\bibnamefont {Pin}}, \
  and\ \bibinfo {author} {\bibfnamefont {M.}~\bibnamefont {Marsili}},\
  }\bibfield  {title} {\emph {\bibinfo {title} {{Assessing the relevance of
  node features for network structure}},\ }}\href {\doibase
  10.1073/pnas.0811511106} {\bibfield  {journal} {\bibinfo  {journal} {Proc
  Natl Acad Sci}\ }\textbf {\bibinfo {volume} {106}},\ \bibinfo {pages} {11433}
  (\bibinfo {year} {2009})}\BibitemShut {NoStop}%
\bibitem [{\citenamefont {Zhao}\ \emph {et~al.}(2011)\citenamefont {Zhao},
  \citenamefont {Halu}, \citenamefont {Severini},\ and\ \citenamefont
  {Bianconi}}]{zhao2011entropy_rate}%
  \BibitemOpen
  \bibfield  {author} {\bibinfo {author} {\bibfnamefont {K.}~\bibnamefont
  {Zhao}}, \bibinfo {author} {\bibfnamefont {A.}~\bibnamefont {Halu}}, \bibinfo
  {author} {\bibfnamefont {S.}~\bibnamefont {Severini}}, \ and\ \bibinfo
  {author} {\bibfnamefont {G.}~\bibnamefont {Bianconi}},\ }\bibfield  {title}
  {\emph {\bibinfo {title} {{Entropy rate of nonequilibrium growing
  networks}},\ }}\href {\doibase 10.1103/PhysRevE.84.066113} {\bibfield
  {journal} {\bibinfo  {journal} {Phys Rev E}\ }\textbf {\bibinfo {volume}
  {84}},\ \bibinfo {pages} {066113} (\bibinfo {year} {2011})}\BibitemShut
  {NoStop}%
\bibitem [{\citenamefont {Peixoto}(2012)}]{peixoto2012entropy}%
  \BibitemOpen
  \bibfield  {author} {\bibinfo {author} {\bibfnamefont {T.~P.}\ \bibnamefont
  {Peixoto}},\ }\bibfield  {title} {\emph {\bibinfo {title} {{Entropy of
  stochastic blockmodel ensembles}},\ }}\href {\doibase
  10.1103/PhysRevE.85.056122} {\bibfield  {journal} {\bibinfo  {journal} {Phys
  Rev E}\ }\textbf {\bibinfo {volume} {85}},\ \bibinfo {pages} {056122}
  (\bibinfo {year} {2012})}\BibitemShut {NoStop}%
\bibitem [{\citenamefont {Peixoto}(2013)}]{peixoto2013parsimonious}%
  \BibitemOpen
  \bibfield  {author} {\bibinfo {author} {\bibfnamefont {T.~P.}\ \bibnamefont
  {Peixoto}},\ }\bibfield  {title} {\emph {\bibinfo {title} {{Parsimonious
  module inference in large networks}},\ }}\href {\doibase
  10.1103/PhysRevLett.110.148701} {\bibfield  {journal} {\bibinfo  {journal}
  {Phys Rev Lett}\ }\textbf {\bibinfo {volume} {110}},\ \bibinfo {pages}
  {148701} (\bibinfo {year} {2013})}\BibitemShut {NoStop}%
\bibitem [{\citenamefont {Anand}\ \emph {et~al.}(2014)\citenamefont {Anand},
  \citenamefont {Krioukov},\ and\ \citenamefont {Bianconi}}]{anand2014entropy}%
  \BibitemOpen
  \bibfield  {author} {\bibinfo {author} {\bibfnamefont {K.}~\bibnamefont
  {Anand}}, \bibinfo {author} {\bibfnamefont {D.}~\bibnamefont {Krioukov}}, \
  and\ \bibinfo {author} {\bibfnamefont {G.}~\bibnamefont {Bianconi}},\
  }\bibfield  {title} {\emph {\bibinfo {title} {{Entropy distribution and
  condensation in random networks with a given degree distribution}},\ }}\href
  {\doibase 10.1103/PhysRevE.89.062807} {\bibfield  {journal} {\bibinfo
  {journal} {Phys Rev E}\ }\textbf {\bibinfo {volume} {89}},\ \bibinfo {pages}
  {062807} (\bibinfo {year} {2014})}\BibitemShut {NoStop}%
\bibitem [{\citenamefont {R{\'{a}}cz}\ and\ \citenamefont
  {Bubeck}(2017)}]{racz2017basic}%
  \BibitemOpen
  \bibfield  {author} {\bibinfo {author} {\bibfnamefont {M.~Z.}\ \bibnamefont
  {R{\'{a}}cz}}\ and\ \bibinfo {author} {\bibfnamefont {S.}~\bibnamefont
  {Bubeck}},\ }\bibfield  {title} {\emph {\bibinfo {title} {{Basic models and
  questions in statistical network analysis}},\ }}\href {\doibase
  10.1214/17-SS117} {\bibfield  {journal} {\bibinfo  {journal} {Stat Surv}\
  }\textbf {\bibinfo {volume} {11}},\ \bibinfo {pages} {1} (\bibinfo {year}
  {2017})}\BibitemShut {NoStop}%
\bibitem [{\citenamefont {Radicchi}\ and\ \citenamefont
  {Castellano}(2018)}]{radicchi2018uncertainty}%
  \BibitemOpen
  \bibfield  {author} {\bibinfo {author} {\bibfnamefont {F.}~\bibnamefont
  {Radicchi}}\ and\ \bibinfo {author} {\bibfnamefont {C.}~\bibnamefont
  {Castellano}},\ }\bibfield  {title} {\emph {\bibinfo {title} {{Uncertainty
  Reduction for Stochastic Processes on Complex Networks}},\ }}\href {\doibase
  10.1103/PhysRevLett.120.198301} {\bibfield  {journal} {\bibinfo  {journal}
  {Phys Rev Lett}\ }\textbf {\bibinfo {volume} {120}},\ \bibinfo {pages}
  {198301} (\bibinfo {year} {2018})}\BibitemShut {NoStop}%
\bibitem [{\citenamefont {Radicchi}\ \emph {et~al.}(2020)\citenamefont
  {Radicchi}, \citenamefont {Krioukov}, \citenamefont {Hartle},\ and\
  \citenamefont {Bianconi}}]{radicchi2020classical}%
  \BibitemOpen
  \bibfield  {author} {\bibinfo {author} {\bibfnamefont {F.}~\bibnamefont
  {Radicchi}}, \bibinfo {author} {\bibfnamefont {D.}~\bibnamefont {Krioukov}},
  \bibinfo {author} {\bibfnamefont {H.}~\bibnamefont {Hartle}}, \ and\ \bibinfo
  {author} {\bibfnamefont {G.}~\bibnamefont {Bianconi}},\ }\bibfield  {title}
  {\emph {\bibinfo {title} {{Classical information theory of networks}},\
  }}\href {\doibase 10.1088/2632-072X/ab9447} {\bibfield  {journal} {\bibinfo
  {journal} {J Phys Complex}\ }\textbf {\bibinfo {volume} {1}},\ \bibinfo
  {pages} {025001} (\bibinfo {year} {2020})}\BibitemShut {NoStop}%
\bibitem [{\citenamefont
  {Bianconi}(2022{\natexlab{a}})}]{bianconi2022statistical}%
  \BibitemOpen
  \bibfield  {author} {\bibinfo {author} {\bibfnamefont {G.}~\bibnamefont
  {Bianconi}},\ }\bibfield  {title} {\emph {\bibinfo {title} {{Statistical
  physics of exchangeable sparse simple networks, multiplex networks, and
  simplicial complexes}},\ }}\href {\doibase 10.1103/PhysRevE.105.034310}
  {\bibfield  {journal} {\bibinfo  {journal} {Phys Rev E}\ }\textbf {\bibinfo
  {volume} {105}},\ \bibinfo {pages} {034310} (\bibinfo {year}
  {2022}{\natexlab{a}})}\BibitemShut {NoStop}%
\bibitem [{\citenamefont {Bianconi}(2022{\natexlab{b}})}]{bianconi2022grand}%
  \BibitemOpen
  \bibfield  {author} {\bibinfo {author} {\bibfnamefont {G.}~\bibnamefont
  {Bianconi}},\ }\bibfield  {title} {\emph {\bibinfo {title} {{Grand Canonical
  Ensembles of Sparse Networks and Bayesian Inference}},\ }}\href {\doibase
  10.3390/e24050633} {\bibfield  {journal} {\bibinfo  {journal} {Entropy}\
  }\textbf {\bibinfo {volume} {24}},\ \bibinfo {pages} {633} (\bibinfo {year}
  {2022}{\natexlab{b}})}\BibitemShut {NoStop}%
\bibitem [{\citenamefont {Cimini}\ \emph {et~al.}(2019)\citenamefont {Cimini},
  \citenamefont {Squartini}, \citenamefont {Saracco}, \citenamefont
  {Garlaschelli}, \citenamefont {Gabrielli},\ and\ \citenamefont
  {Caldarelli}}]{cimini2019statistical}%
  \BibitemOpen
  \bibfield  {author} {\bibinfo {author} {\bibfnamefont {G.}~\bibnamefont
  {Cimini}}, \bibinfo {author} {\bibfnamefont {T.}~\bibnamefont {Squartini}},
  \bibinfo {author} {\bibfnamefont {F.}~\bibnamefont {Saracco}}, \bibinfo
  {author} {\bibfnamefont {D.}~\bibnamefont {Garlaschelli}}, \bibinfo {author}
  {\bibfnamefont {A.}~\bibnamefont {Gabrielli}}, \ and\ \bibinfo {author}
  {\bibfnamefont {G.}~\bibnamefont {Caldarelli}},\ }\bibfield  {title} {\emph
  {\bibinfo {title} {{The statistical physics of real-world networks}},\
  }}\href {\doibase 10.1038/s42254-018-0002-6} {\bibfield  {journal} {\bibinfo
  {journal} {Nat Rev Phys}\ }\textbf {\bibinfo {volume} {1}},\ \bibinfo {pages}
  {58} (\bibinfo {year} {2019})}\BibitemShut {NoStop}%
\bibitem [{\citenamefont {Coutrot}\ \emph {et~al.}(2022)\citenamefont
  {Coutrot}, \citenamefont {Manley}, \citenamefont {Goodroe}, \citenamefont
  {Gahnstrom}, \citenamefont {Filomena}, \citenamefont {Yesiltepe},
  \citenamefont {Dalton}, \citenamefont {Wiener}, \citenamefont
  {H{\"{o}}lscher}, \citenamefont {Hornberger},\ and\ \citenamefont
  {Spiers}}]{coutrot2022entropy}%
  \BibitemOpen
  \bibfield  {author} {\bibinfo {author} {\bibfnamefont {A.}~\bibnamefont
  {Coutrot}}, \bibinfo {author} {\bibfnamefont {E.}~\bibnamefont {Manley}},
  \bibinfo {author} {\bibfnamefont {S.}~\bibnamefont {Goodroe}}, \bibinfo
  {author} {\bibfnamefont {C.}~\bibnamefont {Gahnstrom}}, \bibinfo {author}
  {\bibfnamefont {G.}~\bibnamefont {Filomena}}, \bibinfo {author}
  {\bibfnamefont {D.}~\bibnamefont {Yesiltepe}}, \bibinfo {author}
  {\bibfnamefont {R.~C.}\ \bibnamefont {Dalton}}, \bibinfo {author}
  {\bibfnamefont {J.~M.}\ \bibnamefont {Wiener}}, \bibinfo {author}
  {\bibfnamefont {C.}~\bibnamefont {H{\"{o}}lscher}}, \bibinfo {author}
  {\bibfnamefont {M.}~\bibnamefont {Hornberger}}, \ and\ \bibinfo {author}
  {\bibfnamefont {H.~J.}\ \bibnamefont {Spiers}},\ }\bibfield  {title} {\emph
  {\bibinfo {title} {{Entropy of city street networks linked to future spatial
  navigation ability}},\ }}\href {\doibase 10.1038/s41586-022-04486-7}
  {\bibfield  {journal} {\bibinfo  {journal} {Nature}\ }\textbf {\bibinfo
  {volume} {604}},\ \bibinfo {pages} {104} (\bibinfo {year}
  {2022})}\BibitemShut {NoStop}%
\bibitem [{\citenamefont {Coon}\ \emph {et~al.}(2018)\citenamefont {Coon},
  \citenamefont {Dettmann},\ and\ \citenamefont {Georgiou}}]{coon2018entropy}%
  \BibitemOpen
  \bibfield  {author} {\bibinfo {author} {\bibfnamefont {J.~P.}\ \bibnamefont
  {Coon}}, \bibinfo {author} {\bibfnamefont {C.~P.}\ \bibnamefont {Dettmann}},
  \ and\ \bibinfo {author} {\bibfnamefont {O.}~\bibnamefont {Georgiou}},\
  }\bibfield  {title} {\emph {\bibinfo {title} {{Entropy of spatial network
  ensembles}},\ }}\href {\doibase 10.1103/PhysRevE.97.042319} {\bibfield
  {journal} {\bibinfo  {journal} {Phys Rev E}\ }\textbf {\bibinfo {volume}
  {97}},\ \bibinfo {pages} {042319} (\bibinfo {year} {2018})}\BibitemShut
  {NoStop}%
\bibitem [{\citenamefont {Badiu}\ and\ \citenamefont
  {Coon}(2018)}]{badiu2018distribution}%
  \BibitemOpen
  \bibfield  {author} {\bibinfo {author} {\bibfnamefont {M.-A.}\ \bibnamefont
  {Badiu}}\ and\ \bibinfo {author} {\bibfnamefont {J.~P.}\ \bibnamefont
  {Coon}},\ }in\ \href {\doibase 10.1109/ISIT.2018.8437912} {\emph {\bibinfo
  {booktitle} {2018 IEEE Int Symp Inf Theory}}}\ (\bibinfo  {publisher}
  {IEEE},\ \bibinfo {address} {Vail, CO},\ \bibinfo {year} {2018})\ pp.\
  \bibinfo {pages} {2137--2141}\BibitemShut {NoStop}%
\bibitem [{\citenamefont {Bubeck}\ \emph {et~al.}(2016)\citenamefont {Bubeck},
  \citenamefont {Ding}, \citenamefont {Eldan},\ and\ \citenamefont
  {R{\'{a}}cz}}]{bubeck2016testing}%
  \BibitemOpen
  \bibfield  {author} {\bibinfo {author} {\bibfnamefont {S.}~\bibnamefont
  {Bubeck}}, \bibinfo {author} {\bibfnamefont {J.}~\bibnamefont {Ding}},
  \bibinfo {author} {\bibfnamefont {R.}~\bibnamefont {Eldan}}, \ and\ \bibinfo
  {author} {\bibfnamefont {M.~Z.}\ \bibnamefont {R{\'{a}}cz}},\ }\bibfield
  {title} {\emph {\bibinfo {title} {{Testing for high-dimensional geometry in
  random graphs}},\ }}\href {\doibase 10.1002/rsa.20633} {\bibfield  {journal}
  {\bibinfo  {journal} {Random Struct Algor}\ }\textbf {\bibinfo {volume}
  {49}},\ \bibinfo {pages} {503} (\bibinfo {year} {2016})}\BibitemShut
  {NoStop}%
\bibitem [{\citenamefont {Liu}\ and\ \citenamefont
  {Racz}(2021)}]{liu2021phase}%
  \BibitemOpen
  \bibfield  {author} {\bibinfo {author} {\bibfnamefont {S.}~\bibnamefont
  {Liu}}\ and\ \bibinfo {author} {\bibfnamefont {M.~Z.}\ \bibnamefont {Racz}},\
  }\href@noop {} {\bibfield  {title} {\emph {\bibinfo {title} {{Phase
  transition in noisy high-dimensional random geometric graphs}},\ }}}
  (\bibinfo {year} {2021}),\ \Eprint {http://arxiv.org/abs/2103.15249}
  {arXiv:2103.15249} \BibitemShut {NoStop}%
\bibitem [{\citenamefont {Harary}\ and\ \citenamefont
  {Palmer}(1973)}]{harary1973graphical}%
  \BibitemOpen
  \bibfield  {author} {\bibinfo {author} {\bibfnamefont {F.}~\bibnamefont
  {Harary}}\ and\ \bibinfo {author} {\bibfnamefont {E.~M.}\ \bibnamefont
  {Palmer}},\ }\href {\doibase 10.1016/C2013-0-10826-4} {\emph {\bibinfo
  {title} {{Graphical Enumeration}}}}\ (\bibinfo  {publisher} {Academic
  Press},\ \bibinfo {address} {New York},\ \bibinfo {year} {1973})\BibitemShut
  {NoStop}%
\bibitem [{\citenamefont {Papadopoulos}\ \emph {et~al.}(2018)\citenamefont
  {Papadopoulos}, \citenamefont {Porter}, \citenamefont {Daniels},\ and\
  \citenamefont {Bassett}}]{papadopoulos2018network}%
  \BibitemOpen
  \bibfield  {author} {\bibinfo {author} {\bibfnamefont {L.}~\bibnamefont
  {Papadopoulos}}, \bibinfo {author} {\bibfnamefont {M.~A.}\ \bibnamefont
  {Porter}}, \bibinfo {author} {\bibfnamefont {K.~E.}\ \bibnamefont {Daniels}},
  \ and\ \bibinfo {author} {\bibfnamefont {D.~S.}\ \bibnamefont {Bassett}},\
  }\bibfield  {title} {\emph {\bibinfo {title} {{Network analysis of particles
  and grains}},\ }}\href {\doibase 10.1093/comnet/cny005} {\bibfield  {journal}
  {\bibinfo  {journal} {J Complex Networks}\ }\textbf {\bibinfo {volume} {6}},\
  \bibinfo {pages} {485} (\bibinfo {year} {2018})}\BibitemShut {NoStop}%
\bibitem [{\citenamefont {Kittel}\ and\ \citenamefont
  {Kroemer}(1980)}]{kittel1980thermal}%
  \BibitemOpen
  \bibfield  {author} {\bibinfo {author} {\bibfnamefont {C.}~\bibnamefont
  {Kittel}}\ and\ \bibinfo {author} {\bibfnamefont {H.}~\bibnamefont
  {Kroemer}},\ }\href@noop {} {\emph {\bibinfo {title} {{Thermal Physics}}}},\
  \bibinfo {edition} {2nd}\ ed.\ (\bibinfo  {publisher} {W. H. Freeman and
  Company},\ \bibinfo {address} {New York},\ \bibinfo {year}
  {1980})\BibitemShut {NoStop}%
\bibitem [{\citenamefont {Ford}(2013)}]{ford2013statistical}%
  \BibitemOpen
  \bibfield  {author} {\bibinfo {author} {\bibfnamefont {I.}~\bibnamefont
  {Ford}},\ }\href {\doibase 10.1002/9781118597507} {\emph {\bibinfo {title}
  {{Statistical physics: An entropic approach}}}}\ (\bibinfo  {publisher} {John
  Wiley \& Sons, Ltd},\ \bibinfo {address} {Chichester, UK},\ \bibinfo {year}
  {2013})\BibitemShut {NoStop}%
\bibitem [{\citenamefont {Choi}\ and\ \citenamefont
  {Szpankowski}(2012)}]{choi2012compression}%
  \BibitemOpen
  \bibfield  {author} {\bibinfo {author} {\bibfnamefont {Y.}~\bibnamefont
  {Choi}}\ and\ \bibinfo {author} {\bibfnamefont {W.}~\bibnamefont
  {Szpankowski}},\ }\bibfield  {title} {\emph {\bibinfo {title} {{Compression
  of Graphical Structures: Fundamental Limits, Algorithms, and Experiments}},\
  }}\href {\doibase 10.1109/TIT.2011.2173710} {\bibfield  {journal} {\bibinfo
  {journal} {IEEE Trans Inf Theory}\ }\textbf {\bibinfo {volume} {58}},\
  \bibinfo {pages} {620} (\bibinfo {year} {2012})}\BibitemShut {NoStop}%
\bibitem [{\citenamefont {Kontoyiannis}\ \emph {et~al.}(2020)\citenamefont
  {Kontoyiannis}, \citenamefont {Lim}, \citenamefont {Papakonstantinopoulou},\
  and\ \citenamefont {Szpankowski}}]{kontoyiannis2020compression}%
  \BibitemOpen
  \bibfield  {author} {\bibinfo {author} {\bibfnamefont {I.}~\bibnamefont
  {Kontoyiannis}}, \bibinfo {author} {\bibfnamefont {Y.~H.}\ \bibnamefont
  {Lim}}, \bibinfo {author} {\bibfnamefont {K.}~\bibnamefont
  {Papakonstantinopoulou}}, \ and\ \bibinfo {author} {\bibfnamefont
  {W.}~\bibnamefont {Szpankowski}},\ }\href@noop {} {\bibfield  {title} {\emph
  {\bibinfo {title} {{Compression and Symmetry of Small-World Graphs and
  Structures}},\ }}} (\bibinfo {year} {2020}),\ \Eprint
  {http://arxiv.org/abs/2007.15981} {arXiv:2007.15981} \BibitemShut {NoStop}%
\bibitem [{\citenamefont {Wright}(1974)}]{wright1974graphs}%
  \BibitemOpen
  \bibfield  {author} {\bibinfo {author} {\bibfnamefont {E.~M.}\ \bibnamefont
  {Wright}},\ }\bibfield  {title} {\emph {\bibinfo {title} {{Graphs on
  Unlabelled Nodes with a Large Number of Edges}},\ }}\href {\doibase
  10.1112/plms/s3-28.4.577} {\bibfield  {journal} {\bibinfo  {journal} {Proc
  London Math Soc}\ }\textbf {\bibinfo {volume} {s3-28}},\ \bibinfo {pages}
  {577} (\bibinfo {year} {1974})}\BibitemShut {NoStop}%
\bibitem [{\citenamefont {{\L}uczak}(1991)}]{luczak1991deal}%
  \BibitemOpen
  \bibfield  {author} {\bibinfo {author} {\bibfnamefont {T.}~\bibnamefont
  {{\L}uczak}},\ }\bibfield  {title} {\emph {\bibinfo {title} {{How to deal
  with unlabeled random graphs}},\ }}\href {\doibase 10.1002/jgt.3190150307}
  {\bibfield  {journal} {\bibinfo  {journal} {J Graph Theory}\ }\textbf
  {\bibinfo {volume} {15}},\ \bibinfo {pages} {303} (\bibinfo {year}
  {1991})}\BibitemShut {NoStop}%
\bibitem [{\citenamefont {Bender}\ and\ \citenamefont
  {Canfield}(1978)}]{bender1978asymptotic}%
  \BibitemOpen
  \bibfield  {author} {\bibinfo {author} {\bibfnamefont {E.~A.}\ \bibnamefont
  {Bender}}\ and\ \bibinfo {author} {\bibfnamefont {E.~R.}\ \bibnamefont
  {Canfield}},\ }\bibfield  {title} {\emph {\bibinfo {title} {{The asymptotic
  number of labeled graphs with given degree sequences}},\ }}\href {\doibase
  10.1016/0097-3165(78)90059-6} {\bibfield  {journal} {\bibinfo  {journal} {J
  Comb Theory, Ser A}\ }\textbf {\bibinfo {volume} {24}},\ \bibinfo {pages}
  {296} (\bibinfo {year} {1978})}\BibitemShut {NoStop}%
\bibitem [{\citenamefont {Wormald}(2019)}]{wormald2019asymptotic}%
  \BibitemOpen
  \bibfield  {author} {\bibinfo {author} {\bibfnamefont {N.}~\bibnamefont
  {Wormald}},\ }in\ \href {\doibase 10.1142/9789813272880_0180} {\emph
  {\bibinfo {booktitle} {Proc Int Congr Math (ICM 2018)}}},\ Vol.~\bibinfo
  {volume} {4}\ (\bibinfo  {publisher} {World Scientific},\ \bibinfo {address}
  {Rio de Janeiro},\ \bibinfo {year} {2019})\ pp.\ \bibinfo {pages}
  {3245--3264}\BibitemShut {NoStop}%
\bibitem [{\citenamefont {Wegner}\ and\ \citenamefont
  {Olhede}(2021)}]{wegner2021atomic}%
  \BibitemOpen
  \bibfield  {author} {\bibinfo {author} {\bibfnamefont {A.~E.}\ \bibnamefont
  {Wegner}}\ and\ \bibinfo {author} {\bibfnamefont {S.}~\bibnamefont
  {Olhede}},\ }\bibfield  {title} {\emph {\bibinfo {title} {{Atomic subgraphs
  and the statistical mechanics of networks}},\ }}\href {\doibase
  10.1103/PhysRevE.103.042311} {\bibfield  {journal} {\bibinfo  {journal} {Phys
  Rev E}\ }\textbf {\bibinfo {volume} {103}},\ \bibinfo {pages} {042311}
  (\bibinfo {year} {2021})}\BibitemShut {NoStop}%
\bibitem [{\citenamefont {McKay}\ and\ \citenamefont
  {Wormald}(1984)}]{mckay1984automorphisms}%
  \BibitemOpen
  \bibfield  {author} {\bibinfo {author} {\bibfnamefont {B.~D.}\ \bibnamefont
  {McKay}}\ and\ \bibinfo {author} {\bibfnamefont {N.~C.}\ \bibnamefont
  {Wormald}},\ }\bibfield  {title} {\emph {\bibinfo {title} {{Automorphisms of
  random graphs with specified vertices}},\ }}\href {\doibase
  10.1007/BF02579144} {\bibfield  {journal} {\bibinfo  {journal}
  {Combinatorica}\ }\textbf {\bibinfo {volume} {4}},\ \bibinfo {pages} {325}
  (\bibinfo {year} {1984})}\BibitemShut {NoStop}%
\bibitem [{\citenamefont {Brick}\ \emph {et~al.}(2020)\citenamefont {Brick},
  \citenamefont {Gao},\ and\ \citenamefont {Southwell}}]{brick2020threshold}%
  \BibitemOpen
  \bibfield  {author} {\bibinfo {author} {\bibfnamefont {L.}~\bibnamefont
  {Brick}}, \bibinfo {author} {\bibfnamefont {P.}~\bibnamefont {Gao}}, \ and\
  \bibinfo {author} {\bibfnamefont {A.}~\bibnamefont {Southwell}},\ }\href@noop
  {} {\bibfield  {title} {\emph {\bibinfo {title} {{The threshold of symmetry
  in random graphs with specified degree sequences}},\ }}} (\bibinfo {year}
  {2020}),\ \Eprint {http://arxiv.org/abs/2004.01794} {arXiv:2004.01794}
  \BibitemShut {NoStop}%
\bibitem [{\citenamefont {Cameron}(2013)}]{cameron2013random}%
  \BibitemOpen
  \bibfield  {author} {\bibinfo {author} {\bibfnamefont {P.~J.}\ \bibnamefont
  {Cameron}},\ }\bibinfo {title} {{The Random Graph}},\ in\ \href {\doibase
  10.1007/978-1-4614-7254-4_22} {\emph {\bibinfo {booktitle} {Math Paul Erdős
  II}}}\ (\bibinfo  {publisher} {Springer New York},\ \bibinfo {address} {New
  York, NY},\ \bibinfo {year} {2013})\ pp.\ \bibinfo {pages}
  {353--378}\BibitemShut {NoStop}%
\bibitem [{\citenamefont {Caldarelli}\ \emph {et~al.}(2002)\citenamefont
  {Caldarelli}, \citenamefont {Capocci}, \citenamefont {{De Los Rios}},\ and\
  \citenamefont {Mu{\~{n}}oz}}]{caldarelli2002scale}%
  \BibitemOpen
  \bibfield  {author} {\bibinfo {author} {\bibfnamefont {G.}~\bibnamefont
  {Caldarelli}}, \bibinfo {author} {\bibfnamefont {A.}~\bibnamefont {Capocci}},
  \bibinfo {author} {\bibfnamefont {P.}~\bibnamefont {{De Los Rios}}}, \ and\
  \bibinfo {author} {\bibfnamefont {M.~A.}\ \bibnamefont {Mu{\~{n}}oz}},\
  }\bibfield  {title} {\emph {\bibinfo {title} {{Scale-Free Networks from
  Varying Vertex Intrinsic Fitness}},\ }}\href {\doibase
  10.1103/PhysRevLett.89.258702} {\bibfield  {journal} {\bibinfo  {journal}
  {Phys Rev Lett}\ }\textbf {\bibinfo {volume} {89}},\ \bibinfo {pages}
  {258702} (\bibinfo {year} {2002})}\BibitemShut {NoStop}%
\bibitem [{\citenamefont {S{\"{o}}derberg}(2002)}]{soderberg2002general}%
  \BibitemOpen
  \bibfield  {author} {\bibinfo {author} {\bibfnamefont {B.}~\bibnamefont
  {S{\"{o}}derberg}},\ }\bibfield  {title} {\emph {\bibinfo {title} {{General
  formalism for inhomogeneous random graphs}},\ }}\href {\doibase
  10.1103/PhysRevE.66.066121} {\bibfield  {journal} {\bibinfo  {journal} {Phys
  Rev E}\ }\textbf {\bibinfo {volume} {66}},\ \bibinfo {pages} {066121}
  (\bibinfo {year} {2002})}\BibitemShut {NoStop}%
\bibitem [{\citenamefont {S{\"{o}}derberg}(2003)}]{soderberg2003random}%
  \BibitemOpen
  \bibfield  {author} {\bibinfo {author} {\bibfnamefont {B.}~\bibnamefont
  {S{\"{o}}derberg}},\ }\bibfield  {title} {\emph {\bibinfo {title} {{Random
  graphs with hidden color}},\ }}\href {\doibase 10.1103/PhysRevE.68.015102}
  {\bibfield  {journal} {\bibinfo  {journal} {Phys Rev E}\ }\textbf {\bibinfo
  {volume} {68}},\ \bibinfo {pages} {015102} (\bibinfo {year}
  {2003})}\BibitemShut {NoStop}%
\bibitem [{\citenamefont {Bogu{\~{n}}{\'{a}}}\ and\ \citenamefont
  {Pastor-Satorras}(2003)}]{boguna2003class}%
  \BibitemOpen
  \bibfield  {author} {\bibinfo {author} {\bibfnamefont {M.}~\bibnamefont
  {Bogu{\~{n}}{\'{a}}}}\ and\ \bibinfo {author} {\bibfnamefont
  {R.}~\bibnamefont {Pastor-Satorras}},\ }\bibfield  {title} {\emph {\bibinfo
  {title} {{Class of correlated random networks with hidden variables}},\
  }}\href {\doibase 10.1103/PhysRevE.68.036112} {\bibfield  {journal} {\bibinfo
   {journal} {Phys Rev E}\ }\textbf {\bibinfo {volume} {68}},\ \bibinfo {pages}
  {036112} (\bibinfo {year} {2003})}\BibitemShut {NoStop}%
\bibitem [{\citenamefont {Janson}(2013)}]{janson2013graphons}%
  \BibitemOpen
  \bibfield  {author} {\bibinfo {author} {\bibfnamefont {S.}~\bibnamefont
  {Janson}},\ }\bibfield  {title} {\emph {\bibinfo {title} {{Graphons, cut norm
  and distance, couplings and rearrangements}},\ }}\href@noop {} {\bibfield
  {journal} {\bibinfo  {journal} {NYJM Monogr}\ }\textbf {\bibinfo {volume}
  {4}} (\bibinfo {year} {2013})}\BibitemShut {NoStop}%
\bibitem [{\citenamefont {Kallenberg}(2017)}]{kallenberg2017random}%
  \BibitemOpen
  \bibfield  {author} {\bibinfo {author} {\bibfnamefont {O.}~\bibnamefont
  {Kallenberg}},\ }\href {\doibase 10.1007/978-3-319-41598-7} {\emph {\bibinfo
  {title} {{Random Measures, Theory and Applications}}}},\ \bibinfo {series}
  {Probability Theory and Stochastic Modelling}, Vol.~\bibinfo {volume} {77}\
  (\bibinfo  {publisher} {Springer International Publishing},\ \bibinfo
  {address} {Cham},\ \bibinfo {year} {2017})\BibitemShut {NoStop}%
\bibitem [{\citenamefont {Hanlon}(1982)}]{hanlon1982counting}%
  \BibitemOpen
  \bibfield  {author} {\bibinfo {author} {\bibfnamefont {P.}~\bibnamefont
  {Hanlon}},\ }\bibfield  {title} {\emph {\bibinfo {title} {{Counting Interval
  Graphs}},\ }}\href {\doibase 10.2307/1998705} {\bibfield  {journal} {\bibinfo
   {journal} {Trans Am Math Soc}\ }\textbf {\bibinfo {volume} {272}},\ \bibinfo
  {pages} {383} (\bibinfo {year} {1982})}\BibitemShut {NoStop}%
\bibitem [{\citenamefont {Stanley}\ and\ \citenamefont
  {Fomin}(1999)}]{stanley1999enumerative}%
  \BibitemOpen
  \bibfield  {author} {\bibinfo {author} {\bibfnamefont {R.~P.}\ \bibnamefont
  {Stanley}}\ and\ \bibinfo {author} {\bibfnamefont {S.}~\bibnamefont
  {Fomin}},\ }\href {\doibase 10.1017/CBO9780511609589} {\emph {\bibinfo
  {title} {{Enumerative Combinatorics, Volume 2}}}}\ (\bibinfo  {publisher}
  {Cambridge University Press},\ \bibinfo {address} {Cambridge, UK},\ \bibinfo
  {year} {1999})\BibitemShut {NoStop}%
\bibitem [{\citenamefont {Last}\ and\ \citenamefont
  {Penrose}(2017)}]{last2017lectures}%
  \BibitemOpen
  \bibfield  {author} {\bibinfo {author} {\bibfnamefont {G.}~\bibnamefont
  {Last}}\ and\ \bibinfo {author} {\bibfnamefont {M.}~\bibnamefont {Penrose}},\
  }\href {\doibase 10.1017/9781316104477} {\emph {\bibinfo {title} {{Lectures
  on the Poisson Process}}}}\ (\bibinfo  {publisher} {Cambridge University
  Press},\ \bibinfo {address} {Cambridge, UK},\ \bibinfo {year}
  {2017})\BibitemShut {NoStop}%
\bibitem [{\citenamefont {Orsini}\ \emph {et~al.}(2015)\citenamefont {Orsini},
  \citenamefont {Dankulov}, \citenamefont {Colomer-de Sim{\'{o}}n},
  \citenamefont {Jamakovic}, \citenamefont {Mahadevan}, \citenamefont {Vahdat},
  \citenamefont {Bassler}, \citenamefont {Toroczkai}, \citenamefont
  {Bogu{\~{n}}{\'{a}}}, \citenamefont {Caldarelli}, \citenamefont {Fortunato},\
  and\ \citenamefont {Krioukov}}]{orsini2015quantifying}%
  \BibitemOpen
  \bibfield  {author} {\bibinfo {author} {\bibfnamefont {C.}~\bibnamefont
  {Orsini}}, \bibinfo {author} {\bibfnamefont {M.~M.}\ \bibnamefont
  {Dankulov}}, \bibinfo {author} {\bibfnamefont {P.}~\bibnamefont {Colomer-de
  Sim{\'{o}}n}}, \bibinfo {author} {\bibfnamefont {A.}~\bibnamefont
  {Jamakovic}}, \bibinfo {author} {\bibfnamefont {P.}~\bibnamefont
  {Mahadevan}}, \bibinfo {author} {\bibfnamefont {A.}~\bibnamefont {Vahdat}},
  \bibinfo {author} {\bibfnamefont {K.~E.}\ \bibnamefont {Bassler}}, \bibinfo
  {author} {\bibfnamefont {Z.}~\bibnamefont {Toroczkai}}, \bibinfo {author}
  {\bibfnamefont {M.}~\bibnamefont {Bogu{\~{n}}{\'{a}}}}, \bibinfo {author}
  {\bibfnamefont {G.}~\bibnamefont {Caldarelli}}, \bibinfo {author}
  {\bibfnamefont {S.}~\bibnamefont {Fortunato}}, \ and\ \bibinfo {author}
  {\bibfnamefont {D.}~\bibnamefont {Krioukov}},\ }\bibfield  {title} {\emph
  {\bibinfo {title} {{Quantifying randomness in real networks}},\ }}\href
  {\doibase 10.1038/ncomms9627} {\bibfield  {journal} {\bibinfo  {journal} {Nat
  Commun}\ }\textbf {\bibinfo {volume} {6}},\ \bibinfo {pages} {8627} (\bibinfo
  {year} {2015})}\BibitemShut {NoStop}%
\bibitem [{\citenamefont {Krioukov}\ and\ \citenamefont
  {Ostilli}(2013)}]{krioukov2013duality}%
  \BibitemOpen
  \bibfield  {author} {\bibinfo {author} {\bibfnamefont {D.}~\bibnamefont
  {Krioukov}}\ and\ \bibinfo {author} {\bibfnamefont {M.}~\bibnamefont
  {Ostilli}},\ }\bibfield  {title} {\emph {\bibinfo {title} {{Duality between
  equilibrium and growing networks}},\ }}\href {\doibase
  10.1103/PhysRevE.88.022808} {\bibfield  {journal} {\bibinfo  {journal} {Phys
  Rev E}\ }\textbf {\bibinfo {volume} {88}},\ \bibinfo {pages} {022808}
  (\bibinfo {year} {2013})}\BibitemShut {NoStop}%
\bibitem [{\citenamefont {{\L}uczak}\ \emph {et~al.}(2019)\citenamefont
  {{\L}uczak}, \citenamefont {Magner},\ and\ \citenamefont
  {Szpankowski}}]{luczak2019asymmetry}%
  \BibitemOpen
  \bibfield  {author} {\bibinfo {author} {\bibfnamefont {T.}~\bibnamefont
  {{\L}uczak}}, \bibinfo {author} {\bibfnamefont {A.}~\bibnamefont {Magner}}, \
  and\ \bibinfo {author} {\bibfnamefont {W.}~\bibnamefont {Szpankowski}},\
  }\bibfield  {title} {\emph {\bibinfo {title} {{Asymmetry and structural
  information in preferential attachment graphs}},\ }}\href {\doibase
  10.1002/rsa.20842} {\bibfield  {journal} {\bibinfo  {journal} {Random Struct
  Algor}\ }\textbf {\bibinfo {volume} {55}},\ \bibinfo {pages} {696} (\bibinfo
  {year} {2019})}\BibitemShut {NoStop}%
\bibitem [{\citenamefont {Newman}(2003)}]{newman2003structure}%
  \BibitemOpen
  \bibfield  {author} {\bibinfo {author} {\bibfnamefont {M.~E.~J.}\
  \bibnamefont {Newman}},\ }\bibfield  {title} {\emph {\bibinfo {title} {{The
  Structure and Function of Complex Networks}},\ }}\href {\doibase
  10.1137/S003614450342480} {\bibfield  {journal} {\bibinfo  {journal} {SIAM
  Rev}\ }\textbf {\bibinfo {volume} {45}},\ \bibinfo {pages} {167} (\bibinfo
  {year} {2003})}\BibitemShut {NoStop}%
\bibitem [{\citenamefont {McKay}\ and\ \citenamefont
  {Piperno}(2014)}]{mckay2014practical}%
  \BibitemOpen
  \bibfield  {author} {\bibinfo {author} {\bibfnamefont {B.~D.}\ \bibnamefont
  {McKay}}\ and\ \bibinfo {author} {\bibfnamefont {A.}~\bibnamefont
  {Piperno}},\ }\bibfield  {title} {\emph {\bibinfo {title} {{Practical graph
  isomorphism, II}},\ }}\href {\doibase 10.1016/j.jsc.2013.09.003} {\bibfield
  {journal} {\bibinfo  {journal} {J Symb Comput}\ }\textbf {\bibinfo {volume}
  {60}},\ \bibinfo {pages} {94} (\bibinfo {year} {2014})}\BibitemShut {NoStop}%
\bibitem [{\citenamefont {Wormald}(1987)}]{wormald1987generating}%
  \BibitemOpen
  \bibfield  {author} {\bibinfo {author} {\bibfnamefont {N.~C.}\ \bibnamefont
  {Wormald}},\ }\bibfield  {title} {\emph {\bibinfo {title} {{Generating Random
  Unlabelled Graphs}},\ }}\href {\doibase 10.1137/0216048} {\bibfield
  {journal} {\bibinfo  {journal} {SIAM J Comput}\ }\textbf {\bibinfo {volume}
  {16}},\ \bibinfo {pages} {717} (\bibinfo {year} {1987})}\BibitemShut
  {NoStop}%
\bibitem [{\citenamefont {Hoover}(1979)}]{hoover1979relations}%
  \BibitemOpen
  \bibfield  {author} {\bibinfo {author} {\bibfnamefont {D.~N.}\ \bibnamefont
  {Hoover}},\ }\href@noop {} {\emph {\bibinfo {title} {{Relations on
  Probability Spaces and Arrays of Random Variables}}}},\ \bibinfo {type}
  {Tech. Rep.}\ (\bibinfo  {institution} {Institute for Adanced Study},\
  \bibinfo {address} {Princeton, NJ},\ \bibinfo {year} {1979})\BibitemShut
  {NoStop}%
\bibitem [{\citenamefont {Aldous}(1981)}]{aldous1981representations}%
  \BibitemOpen
  \bibfield  {author} {\bibinfo {author} {\bibfnamefont {D.~J.}\ \bibnamefont
  {Aldous}},\ }\bibfield  {title} {\emph {\bibinfo {title} {{Representations
  for partially exchangeable arrays of random variables}},\ }}\href {\doibase
  10.1016/0047-259X(81)90099-3} {\bibfield  {journal} {\bibinfo  {journal} {J
  Multivar Anal}\ }\textbf {\bibinfo {volume} {11}},\ \bibinfo {pages} {581}
  (\bibinfo {year} {1981})}\BibitemShut {NoStop}%
\end{thebibliography}
